\PassOptionsToPackage{dvipsnames}{xcolor}
\documentclass[preprint]{iacrtrans}
\usepackage[backend=bibtex,style=alphabetic,maxbibnames=99,natbib=true]{biblatex}

\author{Alejandro~Cabrera~Aldaya \and Billy~Bob~Brumley}

\authorrunning{Alejandro Cabrera Aldaya and Billy Bob Brumley}

\index{Aldaya, Alejandro Cabrera}
\index{Brumley, Billy Bob}

\institute{Tampere University, Tampere, Finland\\
\email[alejandro.cabreraaldaya@tuni.fi,billy.brumley@tuni.fi]{alejandro.cabreraaldaya@tuni.fi,billy.brumley@tuni.fi}
}

\bibliography{manuscript_rw,manuscript_ro}

\usepackage[T1]{fontenc}
\usepackage{graphicx}
\usepackage{xspace}
\usepackage{lipsum}
\usepackage{amsmath}
\usepackage{amssymb}
\usepackage{forest}
\usepackage{bm}
\usepackage{fixmath}
\usepackage[linesnumbered,ruled,vlined]{algorithm2e}
\usepackage{url}

\newcommand{\Paragraph}[1]{\paragraph{#1.}}

\newcommand{\footurl}[1]{\footnote{\url{#1}}}

\newcommand{\add}{\emph{addition}\xspace}
\newcommand{\dbl}{\emph{doubling}\xspace}
\newcommand{\Oinf}{\ensuremath{\mathcal{O}}\xspace}

\newcommand{\ie}{i.e.\@\xspace}
\newcommand{\eg}{e.g.\@\xspace}
\newcommand{\cf}{cf.\@\xspace}

\newcommand{\code}[1]{\texttt{#1}\xspace}

\newcommand{\sgxstep}{\code{SGX-Step}}
\newcommand{\graphene}{\code{Graphene-SGX}}

\newcommand{\fl}{\textsc{Flu\-sh+\allowbreak Re\-load}\xspace}

\newcommand{\pp}{\textsc{Pri\-me+\allowbreak Pro\-be}\xspace}

\newcommand{\twodots}{\mathinner {\ldotp \ldotp}}

\newcommand{\lib}[1]{\code{#1}}
\newcommand{\mbedtls}{\lib{mbedTLS}}
\newcommand{\wolfssl}{\lib{wolfSSL}}
\newcommand{\libgcrypt}{\lib{libgcrypt}}

\newcommand{\firststate}{\ensuremath{S^\prime_0}\xspace}
\newcommand{\laststate}{\ensuremath{S^\prime_n}\xspace}
\newcommand{\Lset}{\ensuremath{\boldsymbol{L}}\xspace}

\newcommand{\pt}{\code{PageTracer}}
\newcommand{\cc}{\code{CopyCat}}

\newcommand{\TracerGrind}{\code{TracerGrind}}

\newcommand{\gnuDouble}{\code{gcry\_mpi\_ec\_dup}}
\newcommand{\gnuAdd}{\code{gcry\_mpi\_ec\_add}}

\newcommand{\MBedDouble}{\code{ecp\_double\_jac}}

\newcommand{\wolfDbl}{\code{ecc\_projective\_dbl\_point}}

\newcommand{\circled}[1]{\raisebox{.5pt}{\textcircled{\raisebox{-.9pt} {#1}}}}

\newcommand{\markerPrNF}{1}
\newcommand{\markerPrFP}{2}
\newcommand{\markerPmfL}{A}
\newcommand{\markerCard}{B}
\newcommand{\markerBias}{C}
 
\newcommand{\KEYWORDS}{%
applied cryptography \and
public key cryptography \and
elliptic curve cryptography \and
side-channel analysis \and
online template attacks \and
microarchitecture attacks \and
libgcrypt \and
mbedTLS \and
wolfSSL}

\title{{Online Template Attacks: Revisited}}

\begin{document}

\maketitle

\begin{abstract}
An online template attack (OTA) is a powerful technique previously used
to attack elliptic curve scalar multiplication algorithms.
This attack has only been analyzed in the realm of power consumption and EM side channels,
where the signals leak related to the value being processed.
However, microarchitecture signals have no such feature,
invalidating some assumptions from previous OTA works.

In this paper, we revisit previous OTA descriptions,
proposing a generic framework and evaluation metrics for any side-channel signal.
Our analysis reveals OTA features not previously considered,
increasing its application scenarios and requiring a fresh countermeasure analysis to prevent it.

In this regard, we demonstrate that OTAs can work in the \emph{backward} direction,
allowing to mount an \emph{augmented} projective coordinates attack with respect to
the proposal by Naccache, Smart and Stern\ (Eurocrypt 2004).
This demonstrates that randomizing the initial targeted algorithm state does not prevent the attack as believed in previous works.

We analyze three libraries \libgcrypt, \mbedtls, and \wolfssl
using two microarchitecture side channels.
For the \libgcrypt case, we target its EdDSA implementation using Curve25519 twist curve.
We obtain similar results for \mbedtls and \wolfssl with curve \code{secp256r1}.
For each library, we execute extensive attack instances that are able to recover the complete scalar
in all cases using a single trace.

This work demonstrates that microarchitecture online template attacks are also very powerful
in this scenario, recovering secret information without knowing a leakage model.
This highlights the importance of developing \emph{secure-by-default} implementations,
instead of fix-on-demand ones.
 \keywords{\KEYWORDS{}}
\end{abstract}

\section{Introduction} \label{sec:intro}

Side-channel attacks are a common threat to computing platforms nowadays.
Since the pioneering works by \citet{DBLP:conf/crypto/Kocher96}, several kinds
of leaky channels have been discovered. For instance, execution time, power consumption,
and in the microarchitecture realm cache-timings, sequence of page accesses
\citep{DBLP:conf/crypto/Kocher96,DBLP:conf/crypto/KocherJJ99,Percival05,DBLP:conf/uss/YaromF14,DBLP:conf/sp/XuCP15}.

Several techniques have been proposed to exploit said channels. Among them are template attacks,
that assume the adversary can profile the targeted implementation side-channel signals \citep{DBLP:books/daglib/0017272}.
\citet{DBLP:conf/ches/ChariRR02} introduced template attacks in the context of power consumption side channels,
consisting of three phases:
(i) templates building,
(ii) target trace capturing, and
(iii) template matching.
The template building phase is performed on an attacker-controlled implementation very similar to the targeted one.
During this phase, the attacker profiles leakage by building leakage templates.

Note that this attack description by \citet{DBLP:conf/ches/ChariRR02} considered
templates built \emph{before} capturing the target trace.
Later, \citet{DBLP:conf/wisa/MedwedO08} challenged this order,
with template attacks targeting ECDSA scalar multiplication.
The authors analyzed several scenarios related to this order,
proposing an \emph{on-the-fly} template building \citep[Sect. 5.3]{DBLP:conf/wisa/MedwedO08}.
That is, creating templates \emph{after} capturing the target trace.

Related to this template attack phases order,
\citet{DBLP:conf/indocrypt/BatinaCPST14} proposed what is known in the literature as
Online Template Attacks (OTA),
also building the templates \emph{after} capturing the target trace, similar to \citep{DBLP:conf/wisa/MedwedO08}.
The main difference between \citep{DBLP:conf/wisa/MedwedO08} and \citep{DBLP:conf/indocrypt/BatinaCPST14}
is how templates are constructed.
\citet{DBLP:conf/wisa/MedwedO08} used ``vertical'' power consumption leakage
while \citet{DBLP:conf/indocrypt/BatinaCPST14} used ``horizontal'' leakage
(see \citep{DBLP:conf/icics/ClavierFGRV10} for definitions).
However, despite this difference both approaches agree on the moment when templates are built,
and template attacks that follow this approach are labeled as OTAs.

Creating template traces in advance is feasible when
the number of possible templates to create is small.
For instance, a binary exponentiation algorithm where templates
are used to distinguish a single branch result only requires two templates \citep{DBLP:conf/ches/ChariRR02}.
However, when the number of leaking features to detect is large, \eg coordinates of an elliptic curve point,
the number of different templates could be infeasible to generate in advance.
This scenario is where OTAs enter into play by capturing templates on-demand/\emph{online}
based on a secret guess \citep{DBLP:conf/indocrypt/BatinaCPST14}.

The original OTA technique was proposed and applied in several works using power consumption/EM side channels.
\citet{DBLP:conf/cosade/DugardinPNBDG16} demonstrated a practical OTA on PolarSSL scalar multiplication using EM signals.
Regarding power signals, \citet{otaFourQ} targeted FourQ scalar multiplication and \citet{otaECDSA} instead ECDSA.

\citet{luo2018novel} used the OTA approach, but generated template traces using a leakage model
instead of being captured on a similar device.
This approach requires a leakage model rather than a template device,
but nevertheless adds attack flexibility wrt the original description \citep{DBLP:conf/indocrypt/BatinaCPST14}.

\citet{DBLP:conf/sacrypt/BosFMOS18} analyzed the feasibility of OTAs on the Frodo post-quantum proposal.
The authors employed a power consumption trace emulator for modeling both attack
and template traces instead of real devices \citep{DBLP:conf/uss/McCannOW17}.
It would be interesting to study attack feasibility using such an emulator to gather template traces
while the target trace belongs to a real device, a gap that this paper fills.

One common property in these works is they use \emph{starting algorithm state} (\eg initial elliptic curve point coordinates) as attack input.
This trend is likely motivated by the fact that the OTA technique was originally presented in
this setting, where the starting value of an accumulator is known to the attacker \citep{DBLP:conf/indocrypt/BatinaCPST14}.
However, in this paper we challenge this assumption and show it is not an attack requirement,
considerably expanding its applications.

Regarding microarchitecture side-channel attacks, several template attacks have been proposed in the literature
\citep{DBLP:conf/asiacrypt/BrumleyH09,DBLP:conf/ches/AciicmezBG10,DBLP:conf/uss/GrussSM15,
	DBLP:conf/fc/WeissHS12,chen2019improved,DBLP:conf/ispec/DuL0L15,DBLP:journals/tc/BhattacharyaMBM20}.
\citet{DBLP:conf/asiacrypt/BrumleyH09} developed data cache-timing templates to attack ECDSA scalar multiplication using \pp,
while \citet{DBLP:conf/ches/AciicmezBG10} extended the concept to the L1 instruction cache.
Similarly \citet{DBLP:conf/uss/GrussSM15} showed the feasibility to construct templates from
Last Level Cache timings using \fl to attack AES T-Box implementations.
\citet{DBLP:journals/tc/BhattacharyaMBM20} constructed templates from performance counters related to the branch prediction unit (BPU)
to attack elliptic curve scalar multiplication.

However, regardless of exploited microarchitecture components and applications,
all these works follow the original template attack approach by \citet{DBLP:conf/ches/ChariRR02},
where templates are built \emph{before} capturing the target trace.
Based on extensive literature review, it seems microarchitecture-based OTA related works are a research gap.
Therefore, it remains unknown how OTAs apply in the microarchitecture realm,
especially considering that the original OTA description was motivated by power consumption side-channel
signals that leak differently from microarchitecture ones due to their different natures.
The original OTA technique implicitly assumes information on the side-channel signals that might not be
present in microarchitecture based ones.

In this paper, we start to fill this gap, not only demonstrating the effectiveness of OTAs on
commonly used libraries, but revisiting the original OTA description, demonstrating it is more flexible
than previously believed. This leads to new application scenarios regardless of the exploited side channel.
\autoref{sec:bg} presents background on elliptic curve scalar multiplication algorithms
and microarchitecture side channels.
\autoref{sec:ota} revisits the original OTA description, proposing a generic framework for its analysis.
\autoref{sec:mitigations} analyzes OTA countermeasures considering its previous description and the proposed one.
Later, \autoref{sec:uarch_ota} instantiates the proposed OTA framework in the microarchitecture realm,
evaluating this attack in three software libraries.
\autoref{sec:attacks} presents full end-to-end OTA experiments on these libraries,
capable of recovering the secret in all instances using a single trace.
We present conclusions in \autoref{sec:conclusion}.

Our main contributions are as follows:
(i) we revisit the original OTA concept, revealing properties, application scenarios, and evaluation metrics not considered before;
(ii) we discover that a countermeasure previously proposed to prevent OTAs could be insufficient;
(iii) we propose an \emph{augmented projective coordinates attack} that reduces the number of
required traces from thousands to one, wrt the original attack of \citet{DBLP:conf/eurocrypt/NaccacheSS04};
(iv) we present the first microarchitecture OTA analysis;
(v) we develop a tool to detect OTA-based leakages in software libraries
using different attack vectors;
(vi) we demonstrate practical microarchitecture OTAs on three widely used software libraries.
The first three contributions revisit the original OTA description and are side-channel independent,
whereas the others apply the new methodology to microarchitecture side channels.
 \section{Background}\label{sec:bg}

\subsection{Elliptic curve scalar multiplication}

Scalar multiplication is one of the main operations in ECC.
It computes $P=kG$ for a scalar $k$ and an elliptic curve point $G$,
equivalent to aggregate $k$ times $G$ with itself.
Regarding non-quantum elliptic curve cryptosystems, this operation plays a crucial
role because inverting it (\ie recovering $k$ knowing $P$ and $G$)
requires solving the Elliptic Curve Discrete Logarithm Problem (ECDLP),
considered hard for well-chosen elliptic curves \cite{MR2054891}.

On the other hand, scalar multiplication is the most time-consuming operation
in these cryptosystems.
Among the several approaches to implement it,
performance was initially the main goal.
But after the groundbreaking work on side-channel analysis by \citet{DBLP:conf/crypto/Kocher96}
in \citeyear{DBLP:conf/crypto/Kocher96},
resistance against these attacks is considered a must.

Several approaches exist for computing a scalar multiplication,
for instance: double-and-add, Montgomery ladder, window-based methods, etc.
\citep{DBLP:conf/ches/JoyeY02,cryptoeprint:2004:342,DBLP:conf/ches/Joye07,MR2054891}.
Regardless of their differences, all of them share the property that at every iteration a \emph{state}
is updated based on secret data.
A state can be a single elliptic curve point accumulator (\eg double-and-add)
or a set of them (\eg Montgomery ladder).
This property of scalar multiplication algorithms and the state concept play a crucial role
in the OTA analysis presented in \autoref{sec:ota}.
In this section, we define an abstract scalar multiplication description (\autoref{alg:generic_scalr_mult})
to represent all of them as it fits better for a generic OTA description.
Later during the real-world OTAs in \autoref{sec:attacks},
we instantiate this algorithm using concrete implementations.
\autoref{alg:generic_scalr_mult} consists of four generic operations.

\begin{algorithm}[h]
    \caption{\emph{Generic scalar multiplication}}\label{alg:generic_scalr_mult}
    \DontPrintSemicolon
    \KwIn{Integer $k$ and curve generator $G$}
    \KwOut{$P=kG$}
    \SetKw{KwDownTo}{downto}
    \SetKw{KwIn}{in}
    $K = \code{Encode(}k\code{)}$\\
    $\firststate = \code{Init(}G\code{)}$\\
    \For{$K_i$ \KwIn $K=\{K_1,K_2,\twodots,K_n\}$}{
        $S_i = \code{Select(}S^\prime_{i-1}, K_i\code{)}$\\
        $S^\prime_i = \code{Process(}S_i\code{)}$\\
    }
    $P = \code{Finalize(}\laststate\code{)}$\\
    \Return $P$
\end{algorithm}

\code{Encode}: This operation encodes the scalar $k$ in a list $K=\{K_1,K_2,\twodots,K_n\}$,
where at every algorithm iteration at least one element of $K$ will be processed.
For instance, in the \emph{double-and-add} algorithm, $K$ is the binary representation of $k$.
The encoding defines how many possibilities exist for each $K_i$.
The only requirement for this step is that it can be inverted, \ie it is possible to recover $k$ from $K$.

\code{Init}: This operation initializes the first state \firststate using the point $G$,
as well as performs coordinate conversion and precomputation.

\code{Select}:
This operation defines the state $S_i$ to be processed in the current iteration.
This selection depends on $K_i$ and the previous iteration computed state $S^\prime_{i-1}$.
Sometimes this operation is implemented using conditional branches like in the classic \emph{double-and-add} algorithm,
making it vulnerable to trivial side-channel attacks.
We assume that the implementation of this operation does not leak $K_i$.
This is a common assumption for scalar multiplication algorithms
protected against these trivial attacks (\eg balanced $K_i$-related branches).
Regarding our research, we are more interested in subtle leakages lurking in the ECC hierarchy lower layers,
\eg the finite field implementation.

\code{Process}: This is the most important operation regarding this research.
This step processes the current iteration state $S_i$, generating current iteration \emph{resulting} state $S^\prime_i$.
This operation is composed by elliptic curve point \dbl and \add executions, according to the curve/coordinates formulae.
OTAs aims at identifying which $S_i$ was processed at each iteration, allowing to recover the corresponding $K_i$.
The adversary has freedom to select the \code{Process} operation. For instance, it can be composed
by all point operations in the curve formulae, or only a subset of them.
Additionally, the position of \code{Select} wrt to \code{Process} could vary between implementations,
however adapting the attack to these cases is immediate.

\code{Finalize}: This operation processes the last computed state \laststate just before returning the output point $P$.
For instance, projective to affine coordinates conversion is usually executed here \citep{MR2054891,DBLP:conf/eurocrypt/NaccacheSS04}.

In addition to the scalar multiplication algorithm, there exist several
point coordinate representations that define the formulae employed
for computing point \dbl and \add on a given elliptic curve \citep{MR2054891}.
Regardless of the selected coordinate system,
these operations usually require several modular computations
(\ie additions, multiplications, divisions, etc.) performed on multiprecision integers (\emph{bignum}).
Therefore an ECC implementation consists of several layers,
and eliminating side-channel leakages in all of them is a challenging task.

\subsection{Microarchitecture side-channels}

Several microarchitecture-based side channels have been discovered,
where execution time, cache-timing, and port contention are a few examples
\citep{DBLP:conf/crypto/Kocher96,Percival05,DBLP:conf/uss/YaromF14,DBLP:journals/jce/GeYCH18,DBLP:conf/sp/AldayaBHGT19}.
Despite technique details,
almost all of them aim at exploiting \emph{address-based information leakage}
\citep{book:cryptoeng,DBLP:journals/jce/GeYCH18,DBLP:journals/jhss/Szefer19,DBLP:conf/uss/WeiserZSMMS18}.
That is, a leak produced by \emph{secret-dependent memory accesses}.
When said dependency produces different execution paths,
it is labeled as control-flow leakage,
whereas a data leak exists if a data-memory access is secret-dependent \citep{DBLP:conf/uss/WeiserZSMMS18}.

Another kind of leakage that can exist in a computing platform is \emph{value-based leakage}.
For instance, some CMOS devices leak the Hamming weight of the processed values
through their power consumption \citep{DBLP:conf/crypto/KocherJJ99,DBLP:journals/dt/PoppMO07}.
However, microarchitecture side channels that exploit
value-based leakages are not common at all
\citep{DBLP:conf/sp/CoppensVBS09,DBLP:journals/jce/GeYCH18,DBLP:journals/jhss/Szefer19}.
This difference between power and microarchitecture-based side channels challenges
the application of OTAs in the microarchitecture realm because the original OTA description
inherently assumes that value-based leakage exists in the exploited channel \citep{DBLP:conf/indocrypt/BatinaCPST14}.

The constant-time feature is often used to label a software implementation as side-channel secure.
However, regarding address-based leakages,
a more accurate term is constant-address implementation \citep{DBLP:journals/jce/GeYCH18}.
Nevertheless, we use the term constant-time consistent with the common trend in the literature,
but referring to the latter.

OTAs highlight the need to develop constant-time implementations.
Side-channel secured scalar multiplication algorithms remove secret
dependent branches at its highest level.
However, is it the only layer that needs to be constant-time?
It is common that the lowest \emph{bignum} arithmetic indeed contains branches,
especially when the implementation was inherited from code developed when
side-channel leakages were not a concern.
For instance, OpenSSL, libgcrypt, mbedTLS, wolfSSL are open-source libraries with this property%
\footnote{OpenSSL and wolfSSL are transitioning their \emph{bignum} arithmetic to provide constant-time security at all layers.}.

Address-based memory leakages often exist at different granularities,
and several microarchitecture side channels have been discovered to exploit them.
The zoo of available attacks is diverse with different properties
like threat model, granularity, targeted information, noise level, etc.
The next section presents a generic OTA framework that aims to be applicable to any side channel
(\eg power consumption and microarchitecture signals),
and later in \autoref{sec:uarch_ota} we use it to analyze three libraries employing two microarchitecture ones.
 \section{Reexamining the OTA Technique}\label{sec:ota}

This section revisits the OTA description, proposing a new framework for
its analysis and evaluation, demonstrating some features not considered before.
For this analysis, we use the generic scalar multiplication (\autoref{alg:generic_scalr_mult})
described in \autoref{sec:bg}.

Abstractly, the OTA procedure consists of the following steps.
We discuss differences regarding the original proposal in the following sections.

\begin{enumerate}
	\item Capture one side-channel trace $I$ while the targeted implementation processes some secret $k$.

	\item Split the target trace in iterations $I = \{I_1,I_2,\twodots,I_n\}$,
		  where each $I_i$ corresponds to the processing of state $S_i$ according to $K_i$.
		  Therefore, it is expected $I_i$ corresponds to the \code{Process} operation in \autoref{alg:generic_scalr_mult}.\label{item:iters}

     \item \emph{Extend:} Enumerate every possible $K_i$ and---according to the \code{Select} operation---compute
                          every possible $S_i$ using the previous known state.
                          For each computed $S_i$, capture a side-channel trace
                          corresponding to the execution of \code{Process}$(S_i)$.
                          This produces a template trace $T_{i,j}$ for every possible $K_i$ represented by $j$.\label{item:ota_loop}

	\item \emph{Prune}: Filter out those $T_{i,j}$ that do not \emph{match} $I_i$.\label{item:prune}

	\item For all $T_{i,j}$ surviving the pruning phase,
	 repeat for the next iteration (starting from \autoref{item:ota_loop}), backtracking if multiple matches are found.

     \item \emph{Terminate}: Finish the attack after recovering sufficient $K_i$.
\end{enumerate}

This algorithm follows an \emph{extend-and-prune} approach,
where \autoref{item:ota_loop} \emph{extends} the number of candidates
and \autoref{item:prune} \emph{prunes} unlikely ones.
The OTA idea is to identify which state $S_i$ was processed at an iteration and derive $K_i$ from it.

\Paragraph{Extend}
This step can be performed using different approaches related to
how much control the adversary has over the template implementation.
For instance, which inputs does it accept? Can the adversary modify it?

In this regard, the ideal scenario takes place when the attacker has access to a template implementation
where she can obtain traces of the \code{Process} operation for any state $S_i$.
However, in practice sometimes it is not available due to API limitations.
For instance, in embedded systems it is not common that a device exports an API
for the \code{Process} operation alone, but a high level one for the scalar multiplication,
or even worse, a protocol one (\eg ECDSA signature generation/verification).
In addition, even if the \code{Process} API is available, the state representation
may differ from the targeted one.
For more details on techniques in each scenario, consult
\citep{DBLP:conf/wisa/MedwedO08,DBLP:conf/indocrypt/BatinaCPST14,otaPhD,otaECDSA}.

These limited scenarios are more likely on embedded systems,
in contrast to the microarchitecture realm.
For instance, a common threat model is the adversary and victim share the same computing platform,
attacking a shared cryptography library (\eg cache attacks).
In this scenario, the attacker can use the shared library binary,
or if it is open source she can construct a fork with a more flexible API.
In our practical experiments in \autoref{sec:attacks} we explore this path,
showing this OTA flexibility in the microarchitecture realm.

\Paragraph{Prune}
This phase controls the search tree growth based on a \emph{match score}.
The signal nature determines the method employed to assess if a template trace
matches the targeted one.
Pearson's correlation coefficient has been used in power consumption OTAs
\citep{DBLP:conf/indocrypt/BatinaCPST14,DBLP:conf/cosade/DugardinPNBDG16,otaECDSA},
whereas \citet{DBLP:conf/host/OzgenPB16} explored other classification algorithms.
Regardless of the employed approach, this step should minimize the probability of
pruning the good candidate, while maximizing the probability of pruning incorrect ones.

In this paper, we only consider algorithms that does not prune the correct solution
(\ie $\Pr[\text{false negative}] = 0$).
Dealing with ``false negatives'' requires a highly application-dependent error correction procedure.
For instance, if multiple copies of the target trace can be captured,
this redundancy can help thwart errors, but not all cryptosystems allow this.
We leave the analysis of OTAs combined with error-correction approaches for future work.

\Paragraph{Terminate}
Algorithm termination depends on the attacked cryptosystem and certainty about
the recovered data. Some cryptosystems like ECDSA break when knowing (with
certainty) a small number of bits of the scalars used to generate a set of
signatures \citep{DBLP:journals/dcc/Howgrave-GrahamS01,DBLP:journals/dcc/NguyenS03}.
Therefore, if enough $K_i$ are reliably recovered such that
the number of bits of $k$ that they reveal are sufficient to apply said cryptanalysis,
there is no need to recover the full scalar \citep{DBLP:conf/cosade/DugardinPNBDG16}.

On the other hand, some cryptosystems like the Edwards curve DSA variant (EdDSA) are
designed to prevent such cryptanalysis \citep{DBLP:journals/jce/BernsteinDLSY12}.
For this scenario, OTAs should recover sufficient bits such that solving the ECDLP is feasible,
where naturally recovering the full scalar is also an option.

Partial scalar recovery using OTAs ideally requires a side channel that allows
recovering each $K_i$ with absolute certainty.
Otherwise, it increases the number of iterations to process, expecting that the
pruning removes the incorrect ones \citep{DBLP:conf/cosade/DugardinPNBDG16}.
If a full scalar recovery is desired, the attacker can implement it using
depth-first search, considering that the correct solution will survive every
pruning step and it is more likely that incorrect ones do not. However, if the
pruning phase produces many false positives, the attacker should test each
solution produced by the algorithm until finding the correct one.

\subsection{Attack input and direction}\label{sec:direction}
The OTA concept was presented as a \emph{chosen-input} attack \citep{DBLP:conf/indocrypt/BatinaCPST14}.
However, it is worth highlighting that this requirement is about the template implementation,
not the targeted one.
This distinction is important because it is considered in the OTA literature that
the attacker needs to know the input point to determine the initial state
\firststate and subsequent ones
\citep{DBLP:conf/wisa/MedwedO08,DBLP:conf/indocrypt/BatinaCPST14,
DBLP:conf/cosade/DugardinPNBDG16,DBLP:conf/uss/McCannOW17,luo2018novel,otaECDSA}.

We revisit this claim, discovering that it is not a strict attack requirement.
Instead, we propose the following:
\emph{The OTA technique applies if the adversary knows that a state processed by the target implementation
      belongs to a known set of states with feasible enumeration}.
Note additionally that it is a sufficient condition, not a necessary one (\autoref{sec:requirements} expands).

The initial state \firststate case is covered by previous works.
Moreover, the OTA description given above also applies if the adversary
knows any $S_i$ and starts the attack there.
Such a state could be obtained by a complementary side-channel attack.
We have not found an implementation or previous works with such leakage that allows
recovering an intermediate state.
However, the last state case (\laststate in \autoref{alg:generic_scalr_mult}) requires more attention.

Knowing the last state might seem harmless regarding previous work on OTAs because
it recovers the $K_i$ reproducing the targeted algorithm execution (\ie \emph{forward} direction).
Hence, no state is processed after computing the last state (\cf \autoref{alg:generic_scalr_mult}).
However, we challenge this claim by answering: Could an OTA be executed in the \emph{backward} direction?

Following the OTA description given above,
if the adversary knows the last state $S_n$, she can compute $S_{n-1}$ by inverting the \code{Process} operation.
Depending on the curve formulae, this inversion often involves computing modular roots,
possibly obtaining more than one candidate for $S_{n-1}$ \citep{DBLP:conf/eurocrypt/NaccacheSS04}.
Then the adversary can capture template traces for every computed $S_{n-1}$ and prune those
not matching the observed iteration trace $I_n$.
This will allow determining the processed state and eventually the corresponding $K_i$.
Repeating this process for previous iterations could allow recovering all $K_i$.

This demonstrates OTAs can be applied in the \emph{backward} sense,
reversing the target trace iterations order, \ie $I = \{I_{n}, I_{n-1}, \twodots, I_1\}$
and using the last computed state \laststate as the attack starting state.
This variant could be harder to solve, because each guess for $K_i$ might generate more than one
candidate due to the modular roots, thus increasing the number of candidates per iteration
and the pruning phase must filter out more candidates.
\autoref{sec:attacks} demonstrates this attack in two different scenarios
that recover the scalar using a single trace, showing feasibility in practice.

Very related to this idea is the \emph{projective coordinates attack} proposed by
\citet{DBLP:conf/eurocrypt/NaccacheSS04}.
The authors demonstrated the projective representation of the scalar multiplication output
point could reveal information about the scalar.
The approach is purely algebraic and relies on---when inverting \code{Process}
based on a guess about $K_i$---no modular roots existing, concluding that said $K_i$ is incorrect.
However, due to modular root properties, the search tree explodes very quickly,
hence the number of bits that can be recovered is small \citep{DBLP:conf/eurocrypt/NaccacheSS04,DBLP:journals/tches/AldayaGB20}.

This attack requires knowing the projective coordinates of the output point (\eg last state).
For instance, this can be obtained using a complementary attack:
\citet{DBLP:conf/cosade/MaimutMNT13} used a fault injection attack
while \citet{DBLP:journals/tches/AldayaGB20} used a microarchitecture side channel.
Executing an OTA in the \emph{backward} direction (whereas previous works only considered the \emph{forward} case) could allow recovering
all bits of the scalar using a single trace.
Therefore, it can be considered as an \emph{augmented projective coordinates attack}.

\subsection{Revisited OTA requirements}\label{sec:requirements}

In this section, we revisit the original OTA requirements,
allowing to determine if a scenario could be targeted by an OTA.
Evidently, this does not imply the attack will succeed,
but allows developers to know if their implementation should take OTAs into account.
We define the following OTA requirements:

\begin{itemize}
    \item[] \emph{Distinguisher:} A leak of the \code{Process} implementation can be used as a state distinguisher.

    \item[] \emph{Reproducible:} The adversary has access to a template implementation that
    will process the same data as the target implementation for the same input.
\end{itemize}

The \emph{Distinguisher} requirement has been assumed in previous works due to power consumption side channel properties.
Power consumption signals leak about the values being processed,
therefore this requirement only depends on the signal-to-noise ratio.
However, in the microarchitecture realm, value-based leakages are not common,
therefore the attacker should rely on address-based leakage (\eg non constant-time code).
\autoref{sec:metrics} discusses this requirement and proposes some metrics to evaluate
how well a leak of a \code{Process} implementation can be used as a state distinguisher.

Regarding the \emph{Reproducible} requirement, depending on which input the template implementation accepts,
the attack could be either state- or scalar-based.

\Paragraph{State-based} Previous works on OTAs assume that an attacker knows the first state.
\autoref{sec:direction} extends this to \emph{any} state in \emph{both} forward and backward directions.
This scenario can be generalized even further to the case no state is known,
but the attacker knows a set of states where one of them is the correct one.
Intuitively, if said set can be feasibly enumerated, the adversary can perform the attack for every state in it.

\Paragraph{Scalar-based}
If the template implementation allows executing the same scalar multiplication algorithm
as the target one, the adversary can guess the first processed $K_i$.
For instance, when using a binary algorithm where each $K_i$ represents the bit $i$ of $k$,
the attacker can use the template implementation to capture the traces for $[0]G$ and $[1]G$
and then compare with the target one. If only one matches the target trace, the attacker learns a bit of $k$.
The next iteration builds templates using previously learned information on $k$.
In this scenario, the attacker does not require a known state,
but expects that one of the states processed in template traces corresponds to the target one,
\ie the adversary can \emph{reproduce} the target implementation execution.

The scalar-based approach only works in the forward direction because the processed states depend on previous ones.
During our experiments, we use both approaches to recover the scalar.
As mentioned before, the fulfillment of these requirements does not guarantee attack success.
The next section proposes some metrics to evaluate an implementation regarding OTAs.

\subsection{Evaluation metrics}\label{sec:metrics}

The original OTA paper claims it could recover a full scalar employing
one template trace per key bit \citep{DBLP:conf/indocrypt/BatinaCPST14}.
However, this claim only holds when targeting
a binary scalar multiplication algorithm ($K_i$ is a binary value) and
if the distributions for the \emph{matching scores} for correct and incorrect templates are well-separated.
Therefore, the performance of an OTA depends on the exploited side channel,
its characteristics such as signal-to-noise ratio,
error resilience, etc.
Previous works on OTAs only consider power-based side channels,
making assumptions about the signal that may not hold for other side channels like microarchitecture-based ones.

Following the generic scalar multiplication in \autoref{alg:generic_scalr_mult},
we represent an implementation of the \code{Process} operation using \eqref{eq:process_leak},
where $L_i$ is a side-channel trace resulting from its execution with $S_i$ as input.

\begin{equation}
\label{eq:process_leak}
\code{Process\ensuremath{(S_i)}}  \rightsquigarrow L_i
\end{equation}

Note $L_i$ is equivalent to template traces notation $T_{i,j}$.
However, for the sake of notation simplicity, we rename it as $L_i$ as it fits better
the following analysis where its $K_i$ relation is meaningless.
The objective of each OTA iteration is to detect
which template trace matches with the target one.
Intuitively, the better $L_i$ represents a state $S_i$
the better the attack will perform.

The ideal attacker scenario happens when there is one and only one $L_i$ for every $S_i$
and vice versa (\ie bijective sets).
This implies leakage determinism, that is, every time $S_i$ is processed the same
$L_i$ will be observed.
On the other hand, if the set formed by all possible $L_i$ only has one element,
it is not possible to distinguish any $S_i$ using $L_i$.
Therefore, an implementation with this feature can be considered OTA-safe.

Regarding power consumption side channels, ideal and safe scenarios are not common
due to the channel characteristics.
Power consumption signals contain random noise,
and the ideal scenario can only be achieved if this noise is removed completely,
which is usually not possible in practice \citep{DBLP:books/daglib/0017272,luo2018novel}.
At the same time, a safe scenario (as defined before) is challenging to achieve because
power signals inherently contain value-related leaks (\autoref{sec:bg}),
and preventing those requires specific hardware design \citep{DBLP:books/daglib/0017272}.
Therefore, generally speaking, two different states will inherently generate different signals.
Hence, a power side-channel OTA adversary usually handles scenarios
that lay between these boundary cases.
On the other hand, both ideal and safe scenarios can occur in the microarchitecture realm,
taking advantage of the latter's benefits and suffering the curse of the former
(see countermeasure analysis in \autoref{sec:mitigations}).

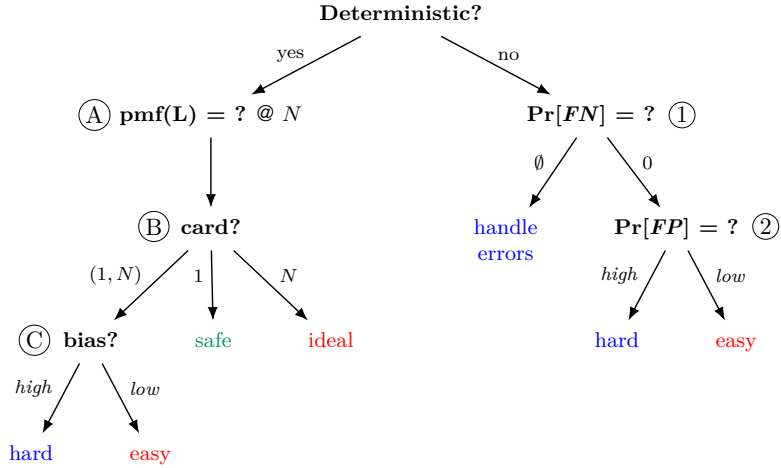
\begin{figure}[h!]
    \newcommand{\safe}{ForestGreen}
    \newcommand{\unsure}{blue}
    \newcommand{\vuln}{red}
    \newcommand{\rootdist}{1.5cm}
    \newcommand{\edgefontsize}{\smaller}
    \newcommand{\ycorrection}{1.5mm}
    \usetikzlibrary{shadows,arrows.meta}
    
    \newcommand{\myat}{{\fontfamily{pag}\selectfont@}\xspace}

    \tikzset{mymarker/.style={draw,solid,circle,inner sep=1pt}}

    \centering
    \resizebox{0.75\columnwidth}{!}{%
    \begin{forest}
    for tree={
            align=center,
            scale=1,
            font=\small,%
            edge={semithick,-Latex},
            for children = {s sep=25pt, l+=20pt}
        },
		[\textbf{Deterministic?},%
        s sep=30mm,%
        [\textbf{pmf(L) = ?} \myat $N$,%
        l=\rootdist,%
        edge label={node[left,midway,xshift=1mm,yshift=1.5mm,font=\edgefontsize]{yes}}%
                [\textbf{card?},
                    [\textbf{bias?}, edge label={node[midway,left,yshift=\ycorrection,font=\edgefontsize]{$(1,N)$}}
                        [hard, text=\unsure, edge label={node[midway,left,yshift=\ycorrection,font=\edgefontsize]{\emph{high}}}]
                        [easy, text=\vuln, edge label={node[midway,right,yshift=\ycorrection,font=\edgefontsize]{\emph{low}}}]
                    ]
                    [safe, text=\safe, edge label={node[midway,left,yshift=\ycorrection,font=\edgefontsize]{$1$}}]
                    [ideal, text=\vuln, edge label={node[midway,right,yshift=\ycorrection,font=\edgefontsize]{$N$}}]
                ]
        ]%
        [{\textbf{Pr[\textit{FN}] = ?}}, l=\rootdist, edge label={node[right,midway,xshift=-1mm,yshift=1.5mm,font=\edgefontsize]{no}}%
            [handle\\errors, text=\unsure, edge label={node[midway,left,yshift=\ycorrection,font=\edgefontsize]{$\emptyset$}}]
            [\textbf{Pr[\textit{FP}] = ?}, edge label={node[midway,right,yshift=\ycorrection,font=\edgefontsize]{$0$}},
                [hard, text=\unsure, edge label={node[midway,left,yshift=\ycorrection,font=\edgefontsize]{\emph{high}}}]
                [easy, text=\vuln, edge label={node[midway,right,yshift=\ycorrection,font=\edgefontsize]{\emph{low}}}]
            ]
        ]
    ]
    \draw(112pt,-40pt)node[mymarker,anchor=west] {\markerPrNF};  %
    \draw(147pt,-87pt)node[mymarker,anchor=west] {\markerPrFP};  %
    \draw(-135pt,-40pt)node[mymarker,anchor=west] {\markerPmfL}; %
    \draw(-110pt,-87pt)node[mymarker,anchor=west] {\markerCard}; %
    \draw(-160pt,-135pt)node[mymarker,anchor=west] {\markerBias};%
	\end{forest}%
    }
	\caption{OTA evaluation flow (attacker perspective).}
	\label{fig:eval_flow}
\end{figure}

We propose some metrics and a procedure allowing
a security auditor or an attacker to evaluate if an implementation could be vulnerable to OTAs.
\autoref{fig:eval_flow} shows a flow diagram to guide the evaluation process.
The first step is to estimate if the targeted implementation is deterministic,
taking into account the considered side channels in the threat model.
Determinism can be estimated by capturing a set of traces with identical inputs
and comparing them.
Ideally this should be done over the entire scalar multiplication algorithm,
to detect which algorithm operations have deterministic behavior.
For instance, non-determinism during the \code{Init} operation is good evidence
there is a state randomization countermeasure in place \citep{DBLP:conf/ches/Coron99},
while a deterministic \code{Init} and \code{Process} could be dangerous.

\Paragraph{Non-deterministic case}
The determinism test defines which main branch of the evaluation flow in \autoref{fig:eval_flow}
should be followed.
If no determinism is observed, the right branch should be taken.
Non-determinism implies the \emph{Reproducible} requirement is not perfectly fulfilled.
Therefore, the template matching algorithm performance should be considered.

For this task, we propose to estimate the probability that said algorithm produces false negatives \circled{\markerPrNF}.
As discussed previously, if it is not zero, the attacker must deal with errors
in the recovery process.
On the other hand, if $\Pr[FN] = 0$ the evaluator knows the solution will remain
in the tree.
Therefore, in this case the \code{Process} leakage
could be \emph{reproduced} somehow.

The last metric for the non-deterministic case allows to estimate the
fulfillment of the \emph{Distinguisher} requirement.
For this purpose, the false positive probability $\Pr[FP]$ can be estimated \circled{\markerPrFP},
\ie probability that the matching algorithm incorrectly classifies a template trace as a match.
This probability defines the number of branches in the solution tree,
therefore it approaches the ideal attacker scenario as the value decreases, and vice versa.
How many false positives the attack can handle depends on the computing resources available to the adversary.

The right branch of \autoref{fig:eval_flow} is likely to occur in power consumption side channels.
Previous OTA works developed attacks with $\Pr[FN] = 0$ and low $\Pr[FP]$ \citep{DBLP:conf/indocrypt/BatinaCPST14,DBLP:conf/cosade/DugardinPNBDG16}.
In general, this branch fits better for noisy side channels.
The left branch of this evaluation flow covers the case where determinism is observed in the
targeted implementation using a particular side channel.
Therefore, the \emph{Reproducible} requirement is perfectly fulfilled in this scenario.

\Paragraph{Deterministic case}
In this scenario, it is possible to evaluate how well a side-channel trace of \code{Process}
can be used as a state \emph{Distinguisher}.
For this task, we propose to estimate the probability mass function ($pmf$)
of the leakages produced by the \code{Process} operation ($L_i$ in \eqref{eq:process_leak}).
We denote this set as \Lset \circled{\markerPmfL}.

The number of possible states is huge, and obtaining an $L_i$ for each of them is infeasible.
Therefore, we estimate the $pmf(\Lset)$ using several ($N$) randomly generated $S_i$,
allowing an estimate of the \emph{cardinality} of \Lset (\ie number of different outcomes),
and how \emph{biased} its distribution is.

\autoref{fig:eval_flow} \circled{\markerCard} shows the conclusions drawn with this estimation.
If the cardinality is one, it means all processed states produced the same leakage.
Therefore, a state cannot be distinguished using the employed side channel, and
the implementation can be considered OTA-safe.
On the contrary, if the number of observed leakages is equal to the number of states used for the
estimation ($N$), it implies an \emph{ideal} attacker scenario.
This means it is very likely an adversary can use the exploited side channel as a perfect state distinguisher.

On the other hand, if the cardinality is between these corner cases, the distribution bias
will determine the computing effort in finding the solution \circled{\markerBias}.
If $pmf(\Lset)$ is highly biased towards one outcome ($L_i$), the number of false positives will
increase, and the search tree grows accordingly.
On the other hand, if no such high bias exists, then solving the problem is easy.
What is considered a high bias depends on attacker computation resources.
During our extensive experiments (detailed later) we recover full 256-bit scalars with bias as high as 62\%
using a desktop workstation.

What could be considered a state \emph{Distinguisher}?
Suppose a state consists of an elliptic curve point,
then very deep in the \emph{bignum} implementation of the targeted implementation
there is a conditional operation that produces two execution branches based on the evenness
of the point coordinate $x$.
Therefore, wlog, assuming a random state, said control-flow leak allows splitting
the state space in two equiprobable halves, \ie such leakage can be used to distinguish if the
processed state contains an even $x$ or not.
This case will produce an equiprobable $pmf(\Lset)$ with two outcomes, yet
even this tiny \emph{state distinguisher} is sufficient to succeed using the OTA technique
(see \autoref{sec:attack_wolfssl} for experiment results).

Note that during an OTA, the adversary is not required to known a model of the exploited leakage,
\ie how a deep \emph{bignum} control-flow leak relates to the processed secret.
Instead, the attacker blindly searches for distinguishable features in the side-channel signals that fulfill OTA requirements.
This blind approach allows to evaluate state-dependent leakages at any layer of the ECC implementation,
no matter how deep they are in the hierarchy.

 \subsection{Mitigation analysis}\label{sec:mitigations}

To prevent OTAs, it should be sufficient to eliminate one of its requirements in the implementation.
For instance, if a call to \code{Process} produces a random signal, it is not possible to \emph{reproduce}
the leakage produced by a given state.
Projective coordinates randomization of the starting state can be used for this purpose,
as proposed in the original OTA paper \citep{DBLP:conf/ches/Coron99,DBLP:conf/indocrypt/BatinaCPST14}.
However, according to the analysis presented in \autoref{sec:direction}, it should
also be applied \emph{after} the scalar multiplication to prevent a \emph{backward} OTA
(\autoref{sec:attack_mbedtls} demonstrates this countermeasure is useless if only executed
at the beginning).
Furthermore, ideally it should be applied to every input state of \code{Process},
thus avoiding a potential OTA based on intermediate states.

Another line of defense is based on thwarting the \emph{Distinguisher} requirement.
That is, prevent a side-channel leak from being used to distinguish the processed state.
Note that this mitigation was not considered in previous works because it is not
easy to achieve in the presence of value-based side channels like power consumption.
However, in the microarchitecture realm, constant-address code should be sufficient
to meet this requirement.
Naturally, this countermeasure will only be effective if it is applied to the entire implementation stack.
 \section{Microarchitecture OTAs}\label{sec:uarch_ota}

In this section, we instantiate the OTA framework from \autoref{sec:ota}
in the microarchitecture realm.
In this context,
a leak can be divided in three groups based on its nature:
(i) executed control-flow,
(ii) data accessed, and
(iii) value processed.
Regarding microarchitecture side channels, the most common leakages are produced by
\emph{secret-dependent memory accesses} (first two cases),
while \emph{value}-based leakages are less common (see \autoref{sec:bg}).

Control-flow based leakages are observed when program execution flow depends on secret data,
\eg due to a conditional instruction result.
While different side channels could be used to exploit such leakages,
in this research we focus our experiments on two approaches, proved very useful to target Intel SGX enclaves.
This scenario is very interesting because SGX technology does not offers protections against
side-channel attacks, delegating such defenses to developers \citep{DBLP:journals/iacr/CostanD16}.
Therefore, analyzing \emph{secured} scalar multiplication implementations in open-source libraries
regarding OTAs is interesting, to assess their resilience to this attack.

Intel SGX aims at providing confidentiality and integrity of software running in Intel microprocessors
even if the OS is under attacker control.
Following this threat model,
the \emph{controlled-channel attack} proposed by \citet{DBLP:conf/sp/XuCP15}
provides access to the sequence of memory pages executed by the victim enclave,
a leakage source with 4 KB granularity that can be used to track the enclave execution
\citep{DBLP:conf/ccs/WangCPZWBTG17,DBLP:conf/ccs/ShindeCNS16,DBLP:conf/uss/BulckWKPS17,DBLP:conf/ccs/WeiserSB18}.
This attack relies on the fact that SGX leaves control of its memory pages to the
untrusted OS. Therefore, an adversarial OS can mark a memory page with SCA relevance as
\emph{non-executable} and monitor it. A triggered page fault indicates the execution
of the monitored page \citep{DBLP:conf/sp/XuCP15}.
Repeating the process for a set of memory pages allows the adversary to track the sequence
of executed memory pages, forming an error-free trace that potentially leaks secret data processed by the enclave.
For the sake of simplicity, we refer to this page tracking attack as \pt.

Recently, \citet{DBLP:journals/corr/abs-2002-08437} proposed the \cc attack
that allows an adversary to glean the number of executed instructions in a tracked memory page.
While it also works at page granularity, it increases the information provided by \pt.
Both attacks can be carried out using the \sgxstep framework proposed by \citet{DBLP:conf/sosp/BulckPS17}.

In this section, we evaluate \mbedtls, \libgcrypt, and \wolfssl scalar multiplication implementations
using the OTA framework proposed in \autoref{sec:ota} regarding \pt and \cc attacks.
This evaluation, in addition to highlighting their vulnerability to OTAs,
extensively compares both attacks and complements the \cc research \citep{DBLP:journals/corr/abs-2002-08437}.

\Paragraph{Threat model}
During the experimental validation of our proposed OTA framework, we employed \pt and \cc attacks.
They share the same SGX threat model:
The adversary has OS privileges and can take advantage of its resources to mount controlled side-channel attacks.
This is a typical threat model for targeting SGX enclaves using side channels
\citep{DBLP:conf/sp/XuCP15,DBLP:conf/sosp/BulckPS17,DBLP:conf/ccs/WeiserSB18,DBLP:journals/tches/AldayaB20}.
In our experiments, we assume an SGX victim application that executes an ECC
scalar multiplication with a secret (\eg EdDSA, ECDSA, and ECDH).

Library selection was not arbitrary. We selected three open-source libraries with
multiprecision integer arithmetic not designed to execute with input-oblivious execution flows.
This selection is interesting, because while these libraries put significant effort in providing
side-channel secure scalar multiplication, usually only the upper layer
receives these security improvements, leaving the \emph{bignum} implementation unattended.
The rationale behind this trend is related to the \emph{fix-on-demand} development process.
At the same time, analyzing how a conditional branch deep in the \emph{bignum} implementation
relates to a secret only processed at the highest layer is usually non-trivial,
as it requires searching for a leakage model---not part of the library development process.
However, OTAs can exploit a deep \emph{bignum} implementation leak without knowing its leakage model.
This highlights the need for \emph{secure-by-default} implementations and the analysis of the selected
libraries regarding OTAs.

We later analyze these libraries, employing the OTA analysis framework proposed in \autoref{sec:ota}.
This demonstrates how an implementation can be evaluated regarding OTA security in practice.
We next introduce tooling that will assist this evaluation process regarding
microarchitecture side channels.
Then in \autoref{sec:attacks}, we employ the results gathered during said OTA evaluation
to develop end-to-end attacks on ECC scalar multiplication implementations in these libraries.

\subsection{Microarchitecture OTA evaluation tool}\label{sec:pmf_attack}

For evaluating OTAs on software libraries using microarchitecture side channels, we developed
a tool that follows step-by-step the evaluation process shown in \autoref{fig:eval_flow}.

We employed \TracerGrind\footurl{https://github.com/SideChannelMarvels/Tracer/tree/master/TracerGrind},
a binary dynamic instrumentation tool developed as part of the Side-Channel Marvels project by \citet{DBLP:conf/ches/BosHMT16}.
This tool patches \code{Valgrind}, allowing to record execution traces of a software binary.
One of its many features is the ability to track a specific address range,
that allows \eg focusing the analysis on a specific shared-library.

\begin{figure}[h]
    \centering
    \includegraphics[width=0.7\linewidth]{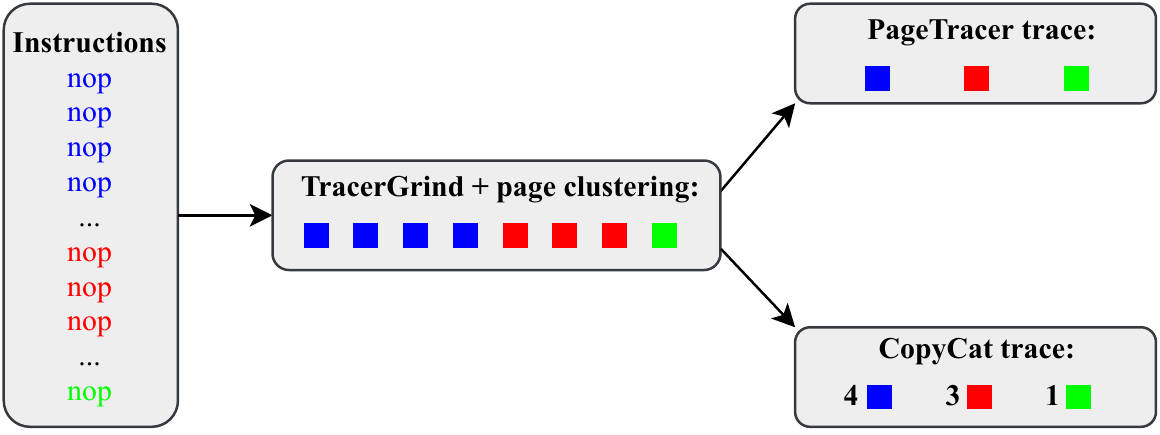}
    \caption{Leakage emulation using \TracerGrind. Colors represent memory pages.}
    \label{fig:tracergrind_emulator}
\end{figure}

The traces recorded with \TracerGrind contain the sequence of accessed addresses (both data and code).
Therefore, it can be used to model a side-channel trace down to instruction granularity.
For instance, \cc can be emulated using \TracerGrind by clearing the 12 least significant bits of
every executed memory address, then run-length-encoding the resulting trace.
This produces a trace that contains the sequence of executed memory pages and the number of instructions
executed within them.
Similarly, a \pt trace can be obtained by removing the instruction count information from a \cc one.
\autoref{fig:tracergrind_emulator} represents this \TracerGrind-based leakage emulation,
highlighting the relationship between executed instructions and the emulated side-channel traces.

\autoref{sec:attacks} empirically validates the accuracy of \TracerGrind regarding \pt,
where it is used to capture the \emph{template traces} during real-world attack instances.
Regarding this evaluation, \TracerGrind allows to emulate a side channel for the \code{Process}
operation on each analyzed library, allowing estimation of the metrics proposed in \autoref{sec:metrics}.
We released a proof-of-concept code that includes this tool and
uses it to emulate a single-trace OTA against \wolfssl \citep{zenodo:2021:ota}.

Naturally, other microarchitecture side-channel signals, like \fl ones at cache-line granularity,
can be modeled using this approach, hence this tool is not side-channel nor SGX-specific.
It is sufficient to encode \TracerGrind output according to the desired leakage.
At the same time, noisy side channels might need a template matching algorithm that supports uncertainty.
For instance, the edit distance can been used in these scenarios
as a template matching metric \citep{DBLP:conf/sp/LiuYGHL15,DBLP:conf/dimva/SchwarzWGMM17}.

\Paragraph{Limitations}
Our tooling aims at detecting OTA-exploitable leaks
considering a given (or modeled) microarchitecture side channel.
Both \pt and \cc work at the memory page boundary (\ie 4kB), therefore the detected leakages depend
on the targeted binary layout.
This means that (in theory) \pt and \cc-vulnerable code might evade detection during evaluation.
For instance with \pt, consider a condition operation with resulting branches executing
in the same page.
In this case, it is not possible to distinguish the executed branches using \pt.
In the opposite case where the target binary has a different layout
such that said branches are in different pages,
the tooling indeed detects the leak using \pt.
Evidently, the probability to detect all OTA-exploitable leakages increases with the granularity,
\ie it is more likely that leaks remain undetected using \pt than \cc.

We designed this tool with an attacker perspective in mind.
Regarding security assessment,
it can be used as a necessary condition to declare an implementation OTA-\emph{safe},
but not as a sufficient criteria.
For this reason, during disclosure to the development teams of the analyzed libraries,
we encouraged generic countermeasures instead of eliminating individual leaks,
also taking into account we only tested them using two side-channel signals
(see \autoref{sec:disclosure} for details).

\subsection{\libgcrypt template implementation}\label{sec:libgcrypt_analysis}

Each library follows its own scalar multiplication approach, therefore the definitions of
\emph{state} and \code{Process} vary.
Scalar multiplication in \libgcrypt v1.8.5 follows the \emph{double-and-add-always} approach
shown in \autoref{alg:double_add}.
It consists of a main loop that iterates over every bit of $k$.
At every iteration, a pair of \dbl and \add operations are executed regardless
of bit $i$ of $k$.
However, even if both operations are executed, $R$ is only updated with the result
of the \add operation if $k_i=1$.
This is ensured by the conditional assignment at \autoref{line:cond} that,
when implemented securely, provides SCA resistance to trivial attacks.

\begin{algorithm}[h]
    \caption{\emph{double-and-add always} scalar multiplication}\label{alg:double_add}
    \DontPrintSemicolon
    \KwIn{Integer $k$ and elliptic curve point $G$}
    \KwOut{$P=kG$}
    \SetKw{KwDownTo}{downto}
    $R = \Oinf$\\
    \For{$i = \lfloor \log_2k \rfloor$ \KwDownTo $0$}{
        $R = 2R$\\
        $T = R + G$\\
        $R =$ \code{cond\_assign($T, R, k_i$)} \label{line:cond} \\
    }
    \Return $R$
\end{algorithm}

Regarding our abstract scalar multiplication description,
the \emph{state} in \autoref{alg:double_add} consists of a single elliptic curve point, $R$.
Similarly, the \code{Process} operation consists of a point \dbl and an \add.
\libgcrypt exports function wrappers for these point operations on Weierstrass and Edwards curves:
\gnuDouble and \gnuAdd respectively.
This allows the attacker to build a \emph{template implementation} to execute both operations
for every input $R$, recording its trace with \TracerGrind.
With \libgcrypt, we focus our research on EdDSA that uses a twisted Edwards
curve birationally equivalent to Curve25519 \citep{DBLP:journals/jce/BernsteinDLSY12}.
The selection of EdDSA is interesting because EdDSA is rarely vulnerable to partial nonce attacks.
Compromising EdDSA requires massive leakage from a single trace, and OTAs fit nicely with this requirement.

Following the evaluation flow in \autoref{fig:eval_flow}, we evaluated the \emph{Reproducible}
requirement to \emph{estimate} if this implementation is deterministic.
For this task, we generated a random curve point and captured 10 independent traces using
our template implementation harness.
We then repeated the experiment for 1000 random points, observing determinism in both \pt and \cc traces in all cases.
Therefore, it is likely that the side-channel trace corresponding to the processing of $R$ can be
\emph{reproduced} with this implementation.
Note that this test demonstrates a fundamental difference between power consumption signals and the employed microarchitecture ones.

According to \autoref{fig:eval_flow}, the next step in a deterministic scenario is to estimate $pmf(\Lset)$.
For this task, we generated 1000 random points, recording their corresponding traces using \TracerGrind.
This experiment resulted in 1000 unique \cc traces,
implying an \emph{ideal} attacker scenario,
and 889 unique \pt traces, which is close to ideal.
Both attack results allow concluding it is very likely
a leakage trace of the \dbl and \add implementations for Edwards curves in \libgcrypt
can be used as a state \emph{distinguisher} for any of these attacks.

Analyzing one of these traces, \libgcrypt executed code in 28 different memory pages
for a single call to the \dbl and \add wrappers for Edwards curves.
We repeated the capture using \TracerGrind but limiting the recording to the
\libgcrypt address space,
hence external calls are not recorded (\eg \code{libc} ones).
The number of executed pages reduces to 17.
An adversary can freely choose their configuration, because during an attack they select which memory pages to track.
However, in our research we are not only interested in determining if \libgcrypt is vulnerable to OTAs,
but analyzing how the leakage behaves considering every memory page combination.
Therefore, as the number of combinations is equal to $2^\ell - 1$ for $\ell$ pages,
we decided to use the limited trace recording to reduce analysis time.

We estimated the $pmf(\Lset)$ for every memory page combination, 131071 in total.
This allows collecting some statistics about OTA performance,
considering \autoref{fig:eval_flow} metrics.
More importantly, it allows pinpointing leakage origins.
\autoref{tab:libgcrypt_metrics_flow_metrics} summarizes how many page combinations were classified
according to the metrics presented in \autoref{sec:metrics}.

\newcommand{\SummaryCaptionLibResult}[2]{{#1} OTA classification for {#2} page combinations.}
\newcommand{\CombCaptionLibResult}[1]{{#1} combination details.}

\begin{table}[h]
\centering
\caption{\SummaryCaptionLibResult{\libgcrypt}{$2^{17}-1$}}
\label{tab:libgcrypt_metrics_flow_metrics}
\begin{tabular}{|c|c|c|c|c|}
    \hline
    \rule[1ex]{0pt}{-0.5ex}                  & \multicolumn{4}{c|}{\textbf{Combination class}} \\
    \hline
    \rule[1ex]{0pt}{-0.5ex} \textbf{Attack}  & \textbf{Ideal} &  \textbf{Easy}  & \textbf{Hard} & \textbf{Safe}  \\
    \hline
    \rule[-1ex]{0pt}{2.5ex} \pt &   0   &  87\%  & 3\%  &   10\%  \\
    \hline
    \rule[-1ex]{0pt}{2.5ex} \cc & 50\%  &  48\%  & 0.8\% & 0.8\%  \\
    \hline
\end{tabular}
\end{table}

Expanding on the number of combinations that could be used to perform an OTA,
\autoref{tab:libgcrypt_comb_details} shows additional results derived from the previous experiment.
The \emph{Insecure} column represents the number of insecure page combinations,
\ie sum of \emph{Ideal} and \emph{Easy} columns in \autoref{tab:libgcrypt_metrics_flow_metrics}.
This means that an attacker can select any \emph{Insecure} page combination to
mount an OTA using these side channels.

\begin{table}
\caption{\CombCaptionLibResult{\libgcrypt}}
\label{tab:libgcrypt_comb_details}
\centering
        \begin{tabular}{|c|c|c|c|c|}
            \hline
            \rule[1ex]{0pt}{-0.5ex} \textbf{Attack}  &   \textbf{Insecure}   & \textbf{Min card} & \textbf{Max bias} & \textbf{Root combs/size} \\
            \hline
            \rule[1.5ex]{0pt}{1ex}     \pt  &      87\%    &    $2$   &   50\%   &       23/2 \\
            \hline
            \rule[1.5ex]{0pt}{1ex}     \cc  &      98\%    &    $7$   &   30\%   &       6/1  \\
            \hline
        \end{tabular}
\end{table}

Among the interesting evaluation data is the \emph{minimum cardinality} that could lead to a successful OTA
and the \emph{maximum bias} observed.
For instance, regarding \pt, at least one insecure page combination exists with only two outcomes in its $pmf(\Lset)$ (cardinality = 2).
At the same time, the maximum bias observed among all combinations is 50\%.
This implies there is at least one combination with a two-cardinality \Lset and equiprobable $pmf$ labeled as insecure.
\autoref{sec:attack_mbedtls} shows the feasibility of attacking this kind of $pmf$.

The last row of \autoref{tab:libgcrypt_comb_details} provides information on the number of \emph{root} page combinations
and their size, \ie number of pages in them.
We define a combination $C$ as root if no smaller combination exists that is a subset of $C$.
For instance, if $C_0 = (P_1)$, $C_1 = (P_1, P_2)$ and $C_2 = (P_1, P_2, P_3)$ are insecure page combinations,
then $C_0$ is a root combination while $C_1$ and $C_2$ are not.
Similarly, if no single page combination were insecure in this example, then $C_1$ would become a root one.
Root combinations can be used to pinpoint where the leakage comes from, especially when they are single-paged ones
like in the \cc case.

Smaller root combinations imply less addresses to be tracked during the attack.
This is not an issue for \pt and \cc due to their noise-free feature.
On the other hand, noisy attacks like \fl will definitely benefit from small root combinations
to decrease noise impact.

\pt has 23 root combinations, all with two pages in them, while \cc has six single-page ones.
Therefore, it could be possible to perform a successful OTA using only two pages for \pt and a single one
for \cc, instead of using all 17 pages involved in our \libgcrypt template implementation.
According to the definition of root, all page combinations are composed by mixing the root ones.
Therefore, in addition to knowing the smaller combinations that could be used to succeed,
it is also interesting to know how many an attacker would need to achieve the maximum cardinality%
\footnote{Here cardinality is used as attack performance, considering that all combinations labeled as insecure
have an \emph{easy} bias, see \autoref{sec:pmf_attack}.}.

\begin{figure}[h]
    \centering
    \includegraphics[width=0.7\linewidth]{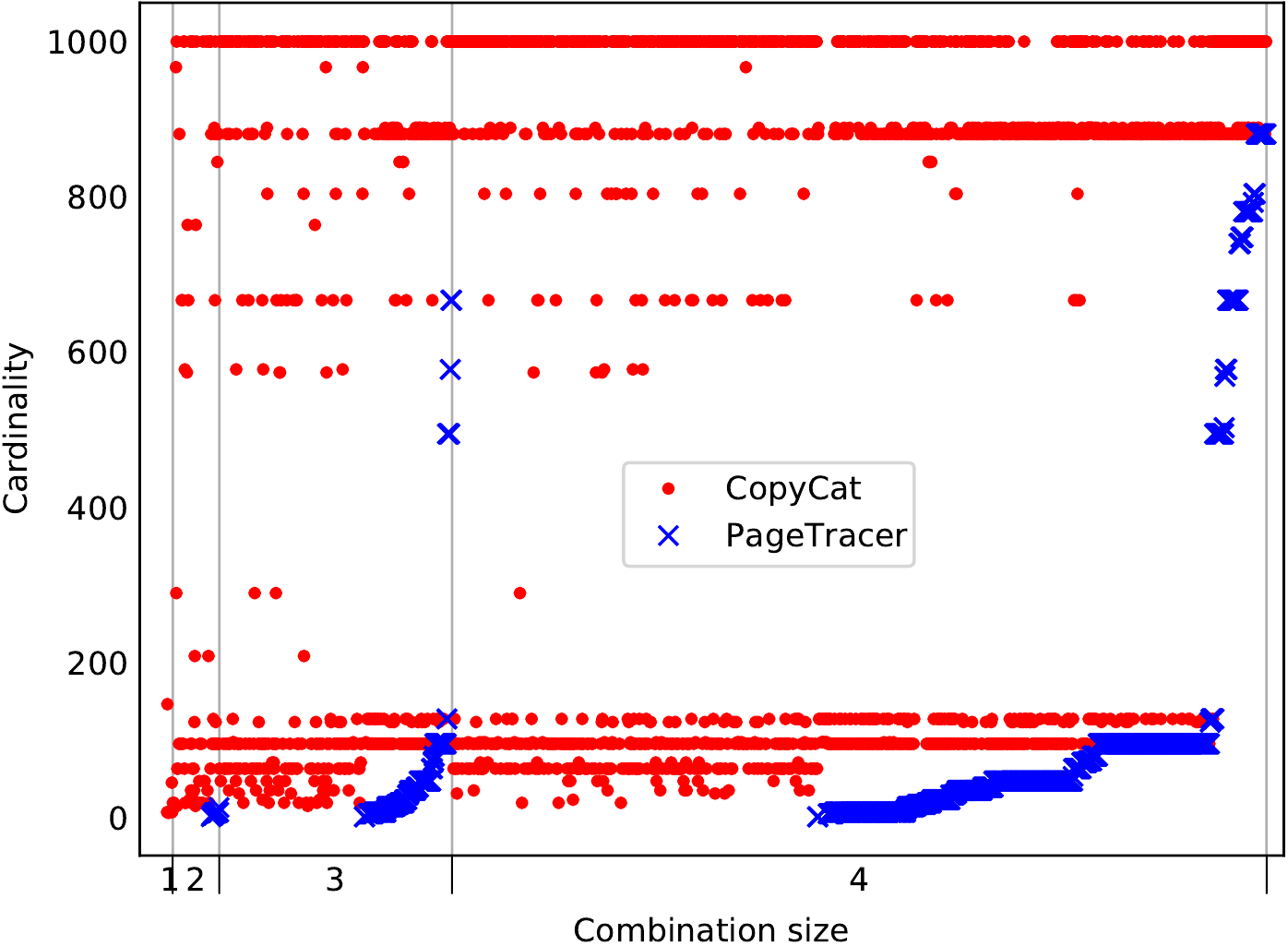}
    \caption{\libgcrypt cardinality vs.\ combination size (partial).}
    \label{fig:libgcrypt_comb_len}
\end{figure}

\autoref{fig:libgcrypt_comb_len} shows how the cardinality progresses as
the number of used memory pages increases from one to four.
Insecure combinations are found starting from two pages and reaching the
maximum cardinality (889 in our experiments) with four pages.
The results for \cc are even better, achieving an \emph{ideal} attack scenario
with only two pages.
Therefore, if an adversary wishes to reduce the number of pages to track,
\eg for noise mitigation or error correction,
both attacks achieved their maximum cardinality even with a reduced set of pages.

\subsection{\mbedtls template implementation}

We analyzed \mbedtls v2.16.3 in the context of the elliptic curve
\code{secp256r1} (\ie NIST P-256).
Elliptic curve computations for this curve use Jacobian projective coordinates.
\autoref{alg:mbedtls_comb} shows a simplified version of the scalar
multiplication algorithm in this library.
It follows a \emph{comb} approach based on the proposal in \citep{cryptoeprint:2004:342}.
This algorithm randomizes the starting value of $R$. However, \autoref{sec:attacks}
expands on how this affects both OTA variants,
surprisingly concluding that it can be ignored for this analysis.

\begin{algorithm}[h]
    \caption{\mbedtls \emph{comb} scalar multiplication}\label{alg:mbedtls_comb}
    \DontPrintSemicolon
    \KwIn{Integer $k$ and elliptic curve point $G$}
    \KwOut{$kG$}
    $K = Encode(k)$\\
    $P = Precompute(G)$\\
    $R = Select(K_1, P)$ \label{line:random}\\
    \For{$K_i \in K : i=[2,n]$}{
        $R = 2R$\\
        $T = Select(K_i, P)$ \label{line:select} \\
        $R = R + T$\\
    }
    \Return $R$
\end{algorithm}

This algorithm encodes the scalar $k$ into a sequence of $K_i$,
where the encoding details are irrelevant because it is invertible.
Thus, if an adversary recovers all $K_i$ she immediately obtains $k$.
The second step precomputes an array $P$ which, for the targeted curve,
contains 32 multiples of $G$.
At each iteration, one point of this array is employed based on $K_i$.
Hence, identifying which point is \emph{selected} at each iteration will reveal $K_i$.
We employ $R$ as \emph{state} in this implementation,
initialized at \autoref{line:random} to an unknown value based on $K_1$.
Still, the attacker knows that $R \in P$ so there are only 32 candidates,
and can start by considering all of them.

To demonstrate the flexibility of OTAs,
for this implementation we chose a \code{Process} operation
composed of only the point \dbl operation, ignoring the leakage produced by \add.
This operation is implemented in function \MBedDouble.
In contrast to the \libgcrypt case, this function is not an exported symbol,
therefore building a template implementation to reach it requires additional effort.
The first strategy we explored was building our own copy of \mbedtls where this symbol is actually
exported, hoping that the symbol table does not significantly change wrt an original (attack) build.
We analyzed both library binaries and the differences were not significant at 4 KB granularity,
therefore we proceeded with this option.

Following the OTA implementation evaluation metrics shown in \autoref{fig:eval_flow},
we used \TracerGrind to assess the determinism of \MBedDouble
using 1000 different points and 10 trials per point.
We deduce that this implementation is likely to be deterministic regarding \pt and \cc.
The execution of \MBedDouble was distributed among 14 memory pages,
meaning 16383 page combinations that could be used to mount an attack.

Similar to our \libgcrypt analysis,
we estimated the $pmf(\Lset)$ for each of these combinations,
and \autoref{tab:mbedtls_comb_details} summarizes the results.
Regarding \cc, all combinations are insecure, with \pt close behind.
Moreover, the number of total \emph{ideal} attacker combinations is very high for both \pt and \cc.
The maximum bias found is 62\% for \pt, yet it is still considered insecure based on our estimations.

\begin{table}[h]
    \caption{\CombCaptionLibResult{\mbedtls}}
    \label{tab:mbedtls_comb_details}
    \centering
        \begin{tabular}{|c|c|c|c|c|c|}
            \hline
            \rule[1ex]{0pt}{-0.5ex} \textbf{Attack}  & \textbf{Ideal} &  \textbf{Insecure}  & \textbf{Min card} & \textbf{Max bias} & \textbf{Root combs/size} \\
            \hline
            \rule[1.5ex]{0pt}{1ex}     \pt           &     84\%       &        99\%         &        $2$        &      62\%         &    63/2 \\
            \hline
            \rule[1.5ex]{0pt}{1ex}     \cc           &     99\%       &       100\%         &        $9$        &      24\%         &    14/1 \\
            \hline
        \end{tabular}
\end{table}

The number of root combinations increases significantly in comparison to \libgcrypt.
Centering the analysis on \cc results,
it is worth highlighting that the number of single-sized root combinations
is equal to the number of memory pages executed by \MBedDouble.
Hence, any page in this set can be used to \emph{distinguish} the processed point.

\begin{figure}[h]
    \centering
    \includegraphics[width=0.7\linewidth]{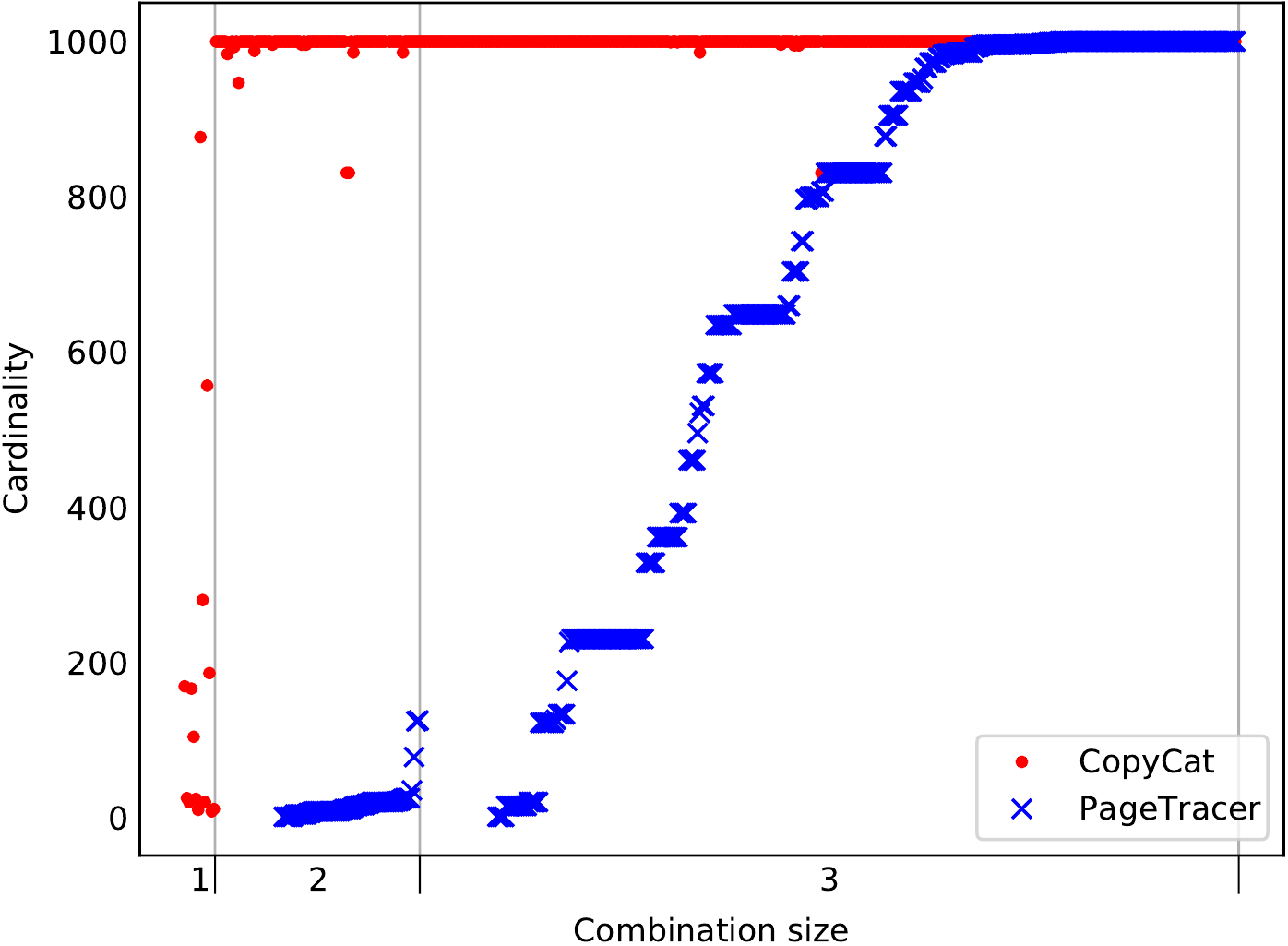}
    \caption{\mbedtls cardinality vs.\ combination size (partial).}
    \label{fig:mbedtls_comb_len}
\end{figure}

\autoref{fig:mbedtls_comb_len} shows the cardinality progression against combination size.
The \emph{ideal} scenario is achieved using \cc for almost every two-page combination,
whereas \pt requires at least three pages to achieve \emph{ideal}.
Both results demonstrate the threat this library faces, especially considering
the high number of small size combinations that achieve the ideal scenario.

\subsection{\wolfssl template implementation}\label{sec:wolfssl_analysis}

We analyzed \wolfssl v4.4.0 with default build options,
including various timing attack countermeasures.
Our analysis focuses on \code{secp256r1}, where
the library uses the Montgomery ladder to compute scalar multiplications (\autoref{alg:mont_ladder}).
The state for this implementation consists of two elliptic curve points $R$ and $S$, initialized to $G$ and $2G$ respectively.
The Montgomery ladder aims at providing side-channel resistance against trivial attacks
by executing a point \add and a \dbl at each iteration, despite the value for bit $i$ of $k$.
However, the arguments for these operations (\ie the state) do depend on $k_i$.

\begin{algorithm}[h]
    \caption{\emph{Montgomery ladder} scalar multiplication}\label{alg:mont_ladder}
    \DontPrintSemicolon
    \KwIn{Positive integer $k$ and elliptic curve point $G$}
    \KwOut{$P=kG$}
    \SetKw{KwDownTo}{downto}
    \SetKwFunction{even}{even}
    $R = G,S = 2G$\\
    \For{$i = \lfloor \log_2(k) \rfloor - 1$ \KwDownTo $0$}{
        \If{$k_i = 0$}{%
            $S = R + S,\enspace R = 2R$%
        }%
        \Else{
            $R = R + S,\enspace S = 2S$%
        }%
    }
    \Return{$R$}
\end{algorithm}

For this implementation, we selected the \dbl operation as our targeted \code{Process},
implemented in function \wolfDbl that is not exported by default.
However, the attacker can build her own version of the library where this symbol is exported, similar to \mbedtls.
Using \TracerGrind, we captured some traces for this function and observed only seven memory pages were executed.
Hence, the number of page combinations is only 127, a considerable reduction wrt to the thousands of \libgcrypt and \mbedtls.

Similar to the previous cases, we estimated the determinism of this \code{Process} implementation,
concluding that it has deterministic leakage for both \pt and \cc.
Following the evaluation flow, we estimated each $pmf(\Lset)$ for each page combination.
\autoref{tab:wolfssl_comb_details} shows the results,
highlighting that the majority of page combinations are insecure using \pt,
and every page combination is insecure using \cc.
For \pt, the maximum observed bias was 52\% with a minimum cardinality of two.
A closer inspection of this leakage revealed it is produced by a modular division by two
which executes an addition before dividing if an intermediate value is odd%
\footurl{https://github.com/wolfSSL/wolfssl/blob/v4.4.0-stable/wolfcrypt/src/ecc.c\#L2179}.

\begin{table}[h]
    \caption{\CombCaptionLibResult{\wolfssl}}
    \label{tab:wolfssl_comb_details}
    \centering
        \begin{tabular}{|c|c|c|c|c|c|}
            \hline
            \rule[1ex]{0pt}{-0.5ex} \textbf{Attack}  & \textbf{Ideal} &  \textbf{Insecure}  & \textbf{Min card} & \textbf{Max bias} & \textbf{Root combs/size} \\
            \hline
            \rule[1.5ex]{0pt}{1ex}         \pt       &         0      &        69\%         &        $2$        &        52\%       &    7/2\\
            \hline
            \rule[1.5ex]{0pt}{1ex}         \cc       &        47\%    &        94\%         &        $7$        &        24\%       &    4/1  \\
            \hline
        \end{tabular}
\end{table}

\begin{figure}[h]
    \centering
    \includegraphics[width=0.7\linewidth]{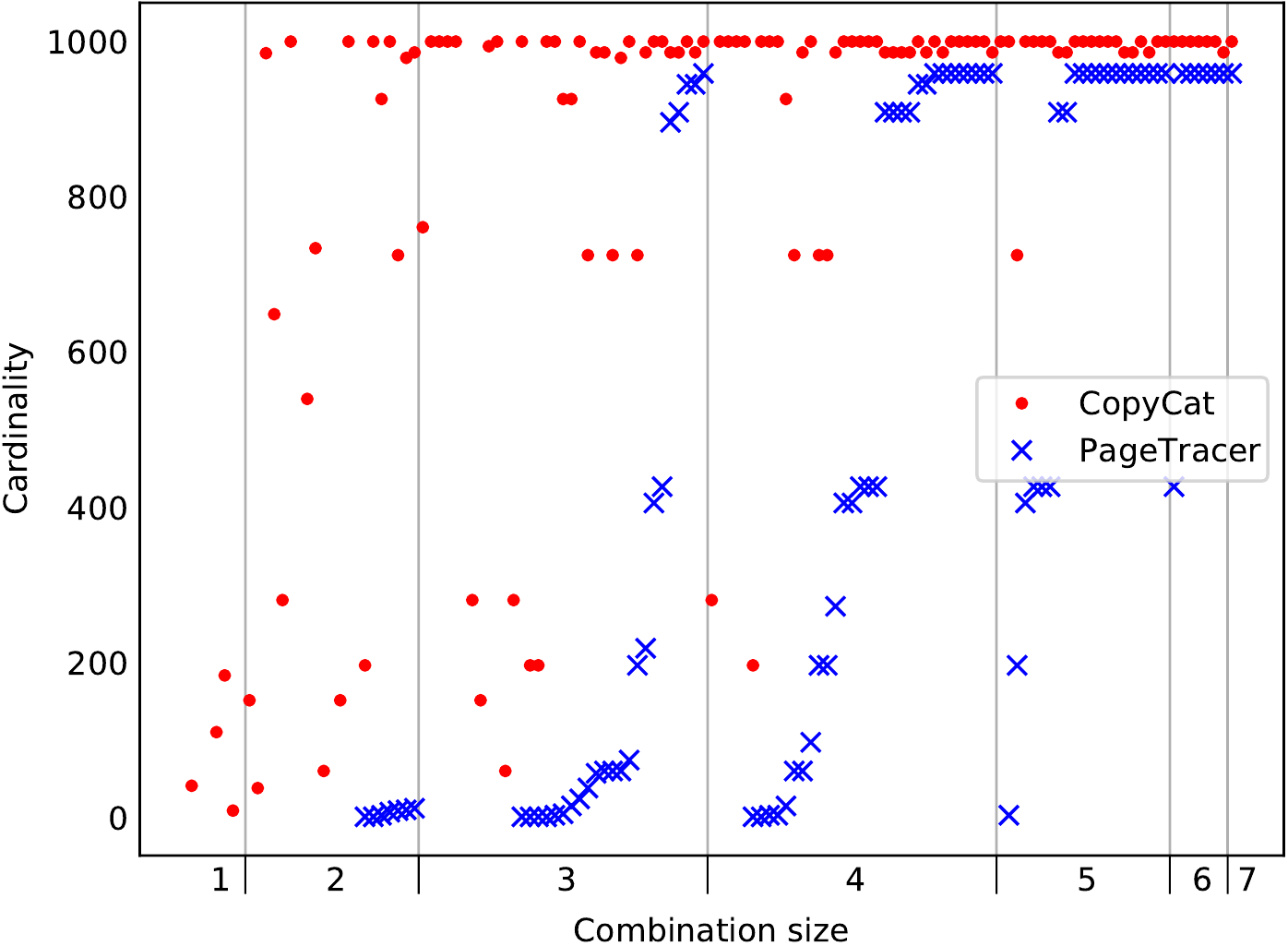}
    \caption{\wolfssl cardinality vs.\ combination size.}
    \label{fig:wolfssl_comb_len}
\end{figure}

\autoref{fig:wolfssl_comb_len} shows how the cardinality progresses with the combination size for \wolfssl.
We reach an \emph{ideal} scenario for two-size page combinations with \cc,
while three pages are required to achieve the maximum cardinality with \pt.
In summary, these results show that even when \wolfssl employs only seven pages
in our targeted \code{Process}, OTAs are possible for many page combinations.

  \section{Real-World Attacks: Evaluating End-to-End OTAs}\label{sec:attacks}

In this section, we develop end-to-end attacks on \libgcrypt, \mbedtls, and \wolfssl scalar multiplication
algorithms using the template implementations described in \autoref{sec:uarch_ota}.
All three attacked libraries share the experiment setup and basic OTA approach.
We consider the threat model described in \autoref{sec:uarch_ota}.

\Paragraph{Experiments setup}
For validating OTAs on the three analyzed libraries, we employed the same environment for capturing the traces.
It consists of a desktop workstation running Ubuntu 18.04.1 LTS on an SGX-capable Intel i7-7700 CPU.
We used the \graphene framework for a straightforward porting of each targeted library to SGX,
requiring no code changes \citep{DBLP:conf/usenix/TsaiPV17}.
We developed \pt-based attacks using the \sgxstep framework \citep{DBLP:conf/sosp/BulckPS17}.
Usually \pt attacks require the adversary to select which pages to track in advance,
based on a known leakage model.
A huge advantage of the $pmf$ analysis performed for each library in previous
sections is that the selection of these pages can be fully automated. That is,
just selecting an insecure page combination, without knowledge of what is
actually executed in those pages.

For all experiments, we compiled the targeted libraries using the latest
versions at the time of writing and the default build options. For each library,
we developed an SGX enclave on top of the \graphene framework. These enclaves
are our attack targets, which instantiate ECC protocols (\libgcrypt) and scalar
multiplication primitives (\mbedtls, \wolfssl) from within their respective libraries.
In all cases, the exploited scalar multiplication
primitives are the default in their libraries for the targeted elliptic curve
and protocol: EdDSA in \libgcrypt, and ECDSA/ECDH in both \mbedtls and \wolfssl.

\Paragraph{Attack implementation}
We implemented OTAs as described in \autoref{sec:bg} using depth-first search
with an early exit when recovering the targeted scalar.
The exit condition varies between library and attack direction.
We followed a \emph{state}-based attack,
therefore the attacker must compute the processed state based on a $K_i$ guess
as explained in \autoref{sec:requirements}.
The recovery code is independent of the targeted $pmf$, therefore we made no optimizations in this regard.

For each library, we selected a $pmf$ and configured \sgxstep to track its corresponding page combination.
After capturing the trace, the recovery code locates the start of the scalar multiplication execution,
then separates its trace in iterations $I = \{I_1,I_2,\twodots,I_n\}$,
where each $I_i$ corresponds to the \code{Process} operation of the implementation.
The OTAs proceed from there.
We give specific details regarding state computation
and scalar multiplication algorithms in the corresponding sections.

\Paragraph{Leakage origins and previous works}
For each attacked library, we pointed to the leaking source code sections to
satisfy reader curiosity. However, we highlight that this information is not
needed by an OTA adversary and was not used in our attacks. One of the
advantages of OTAs is that they can exploit these weaknesses without knowing
those specifics nor the leakage models. During responsible disclosure
(see \autoref{sec:disclosure}), we provided generic leak descriptions and
mitigation approaches.

Previous works on microarchitecture side channels do not cover OTAs. This means
many attacks follow a leakage model-based approach where authors identify a
leaky code path, then find a way to relate it to a secret. A recent example is
\emph{LadderLeak}, where the authors located leaky ECC arithmetic code in older
versions of OpenSSL and developed a leakage model to recover a single bit from
the scalar \citep{DBLP:conf/ccs/AranhaN0TY20}. On the other hand, OTAs remove
the leakage model requirement from the equation. Using tooling presented in
\autoref{sec:uarch_ota} and the evaluation metrics we propose, it is possible to
detect which implementation features (\eg memory pages) can be used to exploit
these leaks, achieving full scalar recovery in many cases.

The leakage origins mentioned in the next sections are only a subset of those
that can be exploited. Enumerating all of them is a time consuming task and
irrelevant to this work. For instance, the number of root page combinations
gives an approximation of this quantity (\eg \mbedtls has 63 for \pt, see
\autoref{tab:mbedtls_comb_details}). Moreover, even those are not exhaustive
because, as previously stated, the tooling was not designed to detect leakages
at all granularities. For the inclined, we suggest tools like the \emph{DATA}
framework which aims at detecting leakages in software binaries using
statistical tools and leakage
models \citep{DBLP:conf/uss/WeiserZSMMS18,DBLP:conf/uss/WeiserSBS20}.
\emph{DATA} is also interesting for our work because its final step uses known
leakage models to assess the severity of a potential leak. However, in many
cases it detects a potential leak but omits information concerning impact. Our
OTA evaluation framework can be used to fill this gap in \emph{DATA} and similar
tools. Finding a potential leak in the ECC scalar multiplication code path using
these leakage detection frameworks, the adversary/evaluator can configure OTA
tooling (see \autoref{sec:uarch_ota}) to estimate the leakage $pmf$. This helps
understand the security impact of these potential leakage points on the
implementation.

\subsection{End-to-end attack on \libgcrypt}\label{sec:attack_libgcrypt}

For an end-to-end attack, we followed the signature generation scenario using
EdDSA with the Curve25519 twist curve (\ie Ed25519).
Generating a signature using Ed25519 involves computing a pseudorandom 512-bit nonce $r$
and computing the scalar multiplication $rG$.
This cryptosystem was designed to avoid \eg lattice
cryptanalysis \citep{DBLP:journals/dcc/Howgrave-GrahamS01} where small
information disclosures on different $r$ break the scheme. For inner details
about this cryptosystem and how $r$ is generated, we refer the reader
to \citep{DBLP:journals/jce/BernsteinDLSY12}. Regarding our research, we aim at
recovering all 512 bits of $r$ that is sufficient to forge signatures, therefore
further details are irrelevant.

\newcommand{\page}[1]{\code{0x#1}}

\begin{figure}[h]
    \centering
    \includegraphics[width=0.7\linewidth]{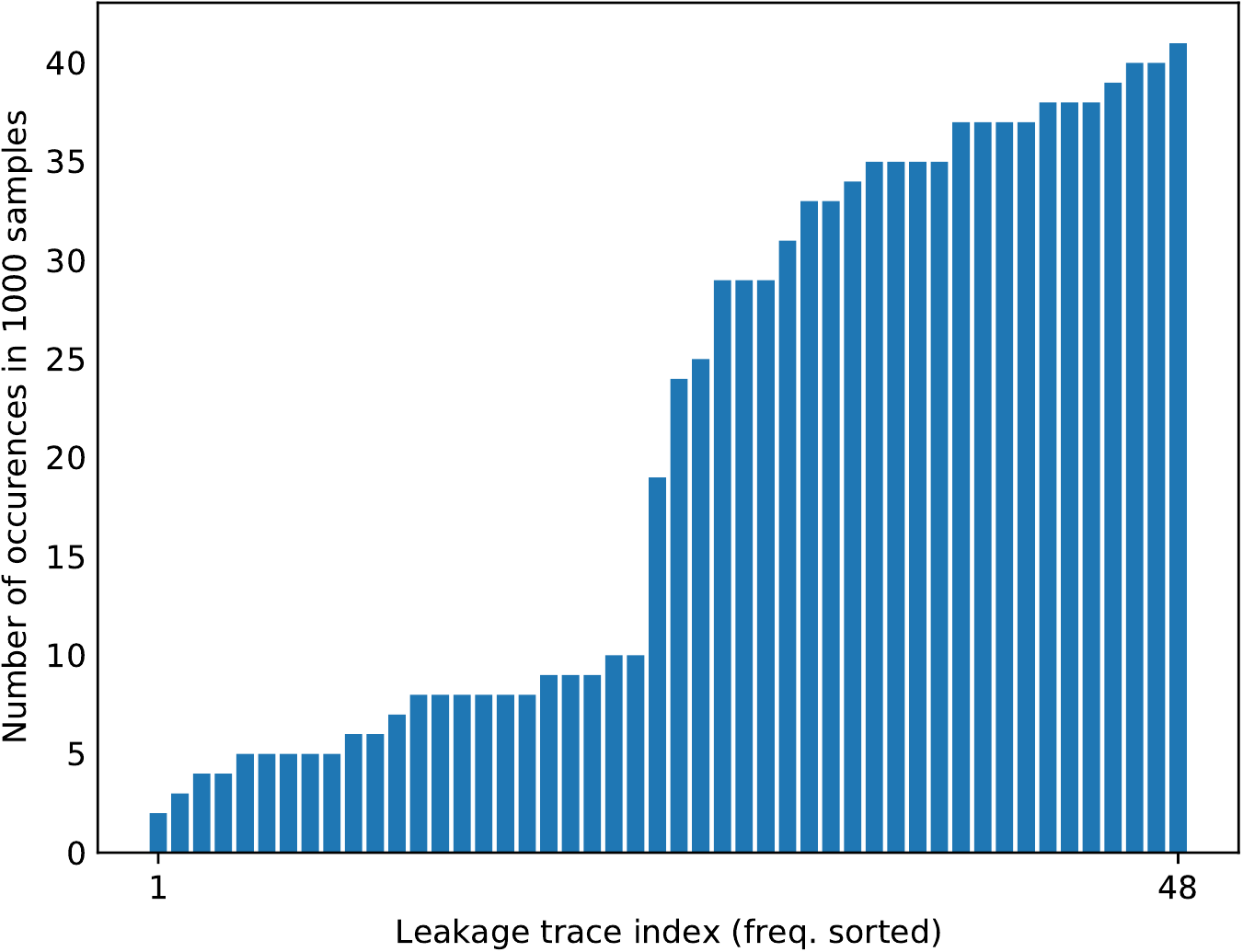}
    \caption{\libgcrypt attack $pmf$.}
    \label{fig:libgcrypt_attack_pmf}
\end{figure}

For this library, we arbitrary selected an insecure (but not ideal) page combination with a $pmf$ of 48 observed outcomes
after 1000 samples (see its $pmf$ in \autoref{fig:libgcrypt_attack_pmf}).
This page combination consists of the following offsets:
\page{d5000}, \page{d6000}, \page{d7000}, \page{d8000}.
Additionally, we used\footnote{No code in this page is executed during scalar
multiplication, not affecting the $pmf$.} offset \page{a3000} to detect when
signature generation starts.

\Paragraph{Forward attack}
Following the \libgcrypt scalar multiplication implementation (\autoref{alg:double_add}),
the adversary must detect which point $R_i$ was processed at each iteration $i$, knowing that
it is initialized to $R_1 = \Oinf$.
This algorithm scans the binary representation of the scalar (\ie $K_i$),
starting from the most significant bit that is always set.
Therefore, at the start of the second iteration, $R_2 = 2R_1 + G = G$, and this is the input to the
\emph{double-and-add}-based \code{Process}.
According to \autoref{alg:double_add}, $R$ is updated at each iteration using \eqref{eq:dbl_add_forward},
hence each $R_i$ depends on $K_{i - 1}$.

\begin{equation}\label{eq:dbl_add_forward}
R_i = 2 R_{i - 1} + G\cdot K_{i - 1}
\end{equation}

Thus, $R_3 = 3G$ if $K_2$ is set and $R_3 = 2G$ otherwise.
Here is where OTAs enter into play by generating template traces for each possible value of $R_3$
(\ie $T_{3,K_2=1}$ and $T_{3,K_2=0}$).
Then eliminating the one that does not match with the target iteration trace (\ie $I_3$).
For this task, we used the template implementation described in \autoref{sec:libgcrypt_analysis}
to generate the required template traces using \TracerGrind.
\autoref{tab:libgcrypt_traces} shows example traces, where the memory pages are encoded
using the characters \{\code{.}, \code{a}, \code{A}, \code{D}\} and the specific assignment is irrelevant.
According to these traces, $I_3$ matches $T_{3,K_2=1}$ and differs from the other:
In this example, the adversary deduces $K_2=1$.
Continuing this process, each $I_i$ and the corresponding template traces can be
used to reveal $K_{i-1}$, eventually disclosing all\footnote{The last $K_i$
remains unknown to an OTA adversary, as there is no related $I_{i+1}$, yet this
binary value is recovered through direct inspection of the two cases along with
public data.} $K_i$.

\newcommand{\traceformat}[1]{\scriptsize{\texttt{#1}}}

\begin{table}[h!]
    \caption{Example traces from \libgcrypt.}
    \label{tab:libgcrypt_traces}
    \centering
    \begin{tabular}{|l|l|} \hline
        \multicolumn{1}{|c|}{\textbf{Name}} & \multicolumn{1}{c|}{\textbf{Trace}} \\ \hline
        $I_3$          & \traceformat{aD.DA.DADA.ADA.ADA.ADA.A.AD.DA.DA.A.AD.DA.D.DA.aA.A.A.A.A.A.A.A.A[..].A.A.A.A.A.A.A.A.A.D}\\ \hline
        $T_{3,K_2=1}$  & \traceformat{aD.DA.DADA.ADA.ADA.ADA.A.AD.DA.DA.A.AD.DA.D.DA.aA.A.A.A.A.A.A.A.A[..].A.A.A.A.A.A.A.A.A.D}\\ \hline
        $T_{3,K_2=0}$  & \traceformat{aD.DA.DA.A.DA.ADA.A.ADADA.ADADA.ADA.DA.aA.A.A.A.A.A.A.A.A.A.A.A.A.A.A.A.A.A.A.D}\\ \hline
    \end{tabular}
\end{table}

We captured 100 \libgcrypt traces corresponding to the generation of Ed25519 signatures.
Using our template implementation based on \TracerGrind, we launched OTAs on them using the
page combination described below,
recovering the processed Ed22519 nonce for each signature in all trials.
The average number of calls to the template implementation was 1038, meaning the
attack is very computationally feasible in practice.

\Paragraph{Backward attack}
This attack assumes the adversary knows the projective coordinates of the last $R_i$ just before converting
the point to affine coordinates.
This requirement can be fulfilled using a side-channel attack on the modular
inversion algorithm in \libgcrypt \citep{DBLP:journals/tches/AldayaGB20}.
The last value of $R$ will be the initial OTA state, thus the adversary reverses the target trace order,
such that the first iteration processed by an OTA will be the last one executed by the algorithm.
In this case, the attacker knows the state resulting from the \code{Process} execution,
\ie $R_i$ in \eqref{eq:dbl_add_forward}.
Therefore, the first step is to compute the \code{Process} input, \ie $R_{i - 1}$ for all possible $K_{i-1}$
using \eqref{eq:rev_dbl_and_add}.

\begin{equation}\label{eq:rev_dbl_and_add}
    R_{i-1} = (R_{i} - G \cdot K_{i - 1}) / 2
\end{equation}

Indeed \eqref{eq:rev_dbl_and_add} implies computing modular roots
\citep{DBLP:conf/eurocrypt/NaccacheSS04,DBLP:journals/tches/AldayaGB20}.
Therefore, multiple $R_{i-1}$ candidates might be obtained for a single pair $(R_i,K_{i-1})$.
This implies more template traces must be captured wrt the \emph{forward} case.

Using this approach, we recovered the scalars processed in all 100 traces,
validating our \emph{augmented projective coordinates attack}.
In this case, the average number of calls to the template implementation was $2851$.
Note that the original projective coordinates attack by \citet{DBLP:conf/eurocrypt/NaccacheSS04}
only recovers a few bits of the scalar, hence does not threaten EdDSA.
On the other hand, our proposed \emph{augmented} version can be used to break this cryptosystem.

\Paragraph{Leakage origin}
The leaks exploited in our attacks come from the \libgcrypt \emph{bignum} implementation.
For instance, the function \code{\_gcry\_mpi\_add} has many conditional branches depending on its inputs%
\footnote{\url{https://github.com/gpg/libgcrypt/blob/libgcrypt-1.8.5/mpi/mpi-add.c\#L88}}.
Each of these branches produce different \pt footprints, hence can be distinguished in a trace.
Another leak is produced by the function \code{ecc\_subm},
where the modular reduction after a subtraction is executed using an add loop%
\footnote{\url{https://github.com/gpg/libgcrypt/blob/libgcrypt-1.8.5/mpi/ec.c\#L290}}.

The advantage of OTAs over ad hoc attacks is that the former can exploit these
leaks without knowing a leakage model. That is, the attacker does not need to
know how a particular branch in these functions leaks information related to the
scalar. Each time these functions are called during an execution of
\code{Process}, they leave leakage footprints related to the data they are
processing. Generally speaking, each of these leakages alone may reveal limited
information, yet their combination composes stronger leakage.

\subsection{End-to-end attack on \mbedtls}\label{sec:attack_mbedtls}

\mbedtls scalar multiplication (\autoref{alg:mbedtls_comb}) has an OTA
countermeasure in place: It randomizes the starting coordinates of $R$ just
after \autoref{line:random}. However, there are at least two scenarios where
OTAs can be applied.
(i) Said point randomization is only executed \emph{before} the scalar
multiplication, therefore it offers protection against a \emph{forward} OTA, but
the \emph{backward} approach is still a threat.
(ii) Said countermeasure only works if an \mbedtls randomization object is
passed as an argument to this function. Regarding ECDSA, this randomization
takes places as expected. Yet we discovered two cases where it fails: When
loading an ECC private key without the public key or in compressed
representation, the library computes the public key on the fly without
initializing the randomization object. This leaves the door open to a
\emph{forward} OTA.

Therefore, the backward case is useful when analyzing protocols like ECDSA or
ECDH, and the forward case when the library loads a private key with missing
(such keys are valid \cite{DBLP:conf/uss/GarciaHTGAB20}) or compressed public
key. Accordingly, we focus our attention on the scalar multiplication primitive
in this library.
The attack procedure is very similar to the \libgcrypt description in
\autoref{sec:attack_libgcrypt}. In the curve \code{secp256r1} case, the scalar
has 256 bits and at each iteration out of 52, the adversary must guess 32
possible $K_i$, due to the windowed feature of \autoref{alg:mbedtls_comb}.

We employed the page combination \page{2f000}, \page{f000}, \page{10000},
\page{36000} that has an \emph{ideal} $pmf$. We used\footnote{No code in these
pages is executed during \MBedDouble, not affecting the $pmf$.} auxiliary page
offsets \page{30000} and \page{31000} to detect the start of the scalar
multiplication routine and the \code{Process} operation (\ie function
\MBedDouble).

We captured 100 traces using \pt against an SGX enclave running scalar multiplication,
attempting to recover the scalar using OTAs in both the forward and backward directions.
The \emph{forward} attack succeeded for all traces, disclosing all\footnote{The
last $K_i$ remains unknown to an OTA adversary, as there is no related
\code{Process} executed after selection, yet this value is recovered through
direct inspection of the 32 cases along with public data.} $K_i$: The number of
calls to the template implementation was $1664$ per attack (\ie 32 templates for
52 iterations). This is due to the \emph{ideal} $pmf$ employed acting as a
perfect state distinguisher.
The \emph{backward} attack instances also succeed for all traces, with an
average number of calls to the template implementation of $1959$. The number
varies between different attacks due to the modular roots involved, generating
additional candidates per each guessed $K_i$ (see \autoref{sec:attack_libgcrypt}
for details). Remarkably, our \emph{backward} OTA bypasses the mitigation
(projective coordinates randomization of the starting state).
A differentiating characteristic compared to the \libgcrypt case is that the
victim executed within an unmodified \mbedtls library, while the template
implementation used a patched one that exported the \MBedDouble symbol.

The leakage in \mbedtls mainly originates from the modular reduction techniques specific to \code{secp256r1}%
\footnote{\url{https://github.com/ARMmbed/mbedtls/blob/mbedtls-2.16.3/library/ecp.c\#L1000}}.
This implementation uses NIST fast reduction techniques, involving several%
\footnote{\url{https://github.com/ARMmbed/mbedtls/blob/mbedtls-2.16.3/library/ecp\_curves.c\#L1010}}
branches%
\footnote{\url{https://github.com/ARMmbed/mbedtls/blob/mbedtls-2.16.3/library/ecp\_curves.c\#L1022}}.
Note that every modular reduction executed during the \code{Process} operation produces leakage.
The combination of all these leaks from branch outcomes produces a perfect state distinguisher, hence an ideal $pmf$.

\subsection{End-to-end attack on \wolfssl}\label{sec:attack_wolfssl}

Finally, we demonstrate the feasibility of OTAs using \pt on \wolfssl, where we
captured 100 traces using \pt against an SGX enclave running scalar multiplication.
We employed the page combination consisting of the offsets
\page{25000}, \page{29000}, \page{2a000}, \page{4b000} that has an \emph{ideal}
$pmf$. We additionally used\footnote{No code in this page is executed during
\wolfDbl, not affecting the $pmf$.} \page{2c000} to detect the start of the
scalar multiplication routine.

We executed only \emph{forward} OTAs against \wolfssl, resulting in full key
recovery in all trials, requiring 512 calls to the template implementation in
all cases. Note that an ideal $pmf$ and a binary scalar multiplication algorithm
like the Montgomery ladder allow reducing this number to only 256 (\ie one call
per bit). However, as stated at the beginning of this section, we implemented
our recovery algorithm to always be oblivious to the $pmf$ to retain generality.

In addition to the previous $pmf$, we attacked this implementation using other
approaches.
(i) We considered a two-cardinality equiprobable $pmf$, and executed this attack
for three traces, recovering the full scalar in all trials. Naturally, the
number of calls to the template implementation increases, and also the number of
solutions to test. The pairs for these values for the three instances were
$(5780, 10)$, $(19454, 76)$, and $(30640, 177)$, all practical attacks.
(ii) We considered the \emph{scalar}-based OTA theoretically presented in
\autoref{sec:requirements}. Instead of guessing the state, this approach assumes
the adversary has access to a template implementation allowing chosen-input
scalars. For this scenario, we used 100 traces and the same ideal page
combination employed before, indeed recovering the full scalar in all trials.
In addition, we released a proof-of-concept tooling that simulates this OTA approach
against this library \citep{zenodo:2021:ota}.

The leaks exploited in our \wolfssl OTAs come in different flavors.
For instance, the \emph{doubling} function \code{ecc\_projective\_dbl\_point}
contains several branches related to its inputs%
\footnote{\url{https://github.com/wolfSSL/wolfssl/blob/v4.4.0-stable/wolfcrypt/src/ecc.c\#L1932}}.
The modular division by two is straightforwardly implemented using a branch that first checks if the input is odd%
\footnote{\url{https://github.com/wolfSSL/wolfssl/blob/v4.4.0-stable/wolfcrypt/src/ecc.c\#L2179}}.
Modular additions%
\footnote{\url{https://github.com/wolfSSL/wolfssl/blob/v4.4.0-stable/wolfcrypt/src/ecc.c\#L2133}}
and subtractions%
\footnote{\url{https://github.com/wolfSSL/wolfssl/blob/v4.4.0-stable/wolfcrypt/src/ecc.c\#L2201}}
are handled in similar ways.
Additionally, the functions \code{fp\_sub}%
\footnote{\url{https://github.com/wolfSSL/wolfssl/blob/v4.4.0-stable/wolfcrypt/src/tfm.c\#L166}}
and \code{fp\_montgomery\_reduce}%
\footnote{\url{https://github.com/wolfSSL/wolfssl/blob/v4.4.0-stable/wolfcrypt/src/tfm.c\#L3035}}
have \pt-distinguishable branches in their implementations.

As commented at the start of this section, the attacker does not need to know where the
leakage originates, rather the page combination that leads to a successful
attack. This can be identified during a template implementation evaluation, as
explained in \autoref{sec:uarch_ota}. This way, OTAs abstract the exploited
leakage model from the adversary, resulting in very powerful attacks.
\section{Conclusion} \label{sec:conclusion}

Previous works related to the OTA technique only considered part of its
potential. In this paper, we revisited that description, proposing a framework
and evaluation metrics to detect if an implementation is vulnerable to OTAs.
Additionally, we demonstrated that OTAs can also work in the \emph{backward}
direction, a case not considered before. This shows that randomizing the initial
state of the targeted algorithm does not blanketly prevent OTAs, as previously
believed. In this regard, an \emph{augmented projective coordinate attack} is
one example of a \emph{backward} OTA because it can recover the entire scalar
using a single trace. This is in contrast to the thousands needed by the
original projective coordinates attack by \citet{DBLP:conf/eurocrypt/NaccacheSS04}.

The three analyzed libraries \libgcrypt, \mbedtls, and \wolfssl have many leaky
points that can be exploited using OTAs. We demonstrated practical attacks for
the three libraries, in all cases recovering the full scalar by employing a single
trace using a microarchitecture side channel after extensive experiments.
In the microarchitecture realm, it is possible to have an \emph{ideal} attacker
scenario as demonstrated for the analyzed libraries. At the same time, it is
also possible to achieve \emph{safe} ones if the implementation follows a
constant-address approach. These scenarios are not common at all in the power
consumption case, where the original OTA technique was proposed.

Our tool proposed to detect OTA vulnerabilities is not exhaustive, therefore
there could be additional exploitable paths. At the same time, its idea serves
as a starting point to develop a leakage assessment tool for address-based side
channels. Such a tool is ideally able to detect any OTA vulnerability in the
hierarchy of a cryptosystem implementation, certainly not restricted to ECC.

In conclusion, OTAs can exploit non-trivial input-dependent execution flows
without knowing the leakage model, highlighting the need for
\emph{secure-by-default} implementations.

\Paragraph{Acknowledgments}
This project has received funding from the European Research Council (ERC) under
the European Union's Horizon 2020 research and innovation programme (grant
agreement No 804476).

Supported in part by CSIC's i-LINK+ 2019 ``Advancing in cybersecurity technologies''
(Ref.\ LINKA20216).
 
\printbibliography

\clearpage
\appendix
\section{Disclosure and library responses}\label{sec:disclosure}

We contacted the development teams of the analyzed libraries to disclose the leakage sources,
proposing countermeasures according to the analysis in \autoref{sec:mitigations}.
We did not focus the disclosure nor our work on specific vulnerabilities,
but instead on a generic countermeasure approach.
Therefore, the proposed countermeasures were independent of the exploited code path
we used in our practical evaluation (\autoref{sec:attacks}).

During disclosure, we highlighted the strengths of the projective coordinates randomization countermeasure
over constant-time code (\ie address-independent code):
\begin{enumerate}

\item Constant-time code only prevents address-based leakage OTAs (likely sufficient to prevent microarchitecture attacks),
but does not thwart value-base leakage OTAs like power consumption.

\item The experiments in this work employed \pt and \cc traces,
therefore even an implementation with a safe $pmf$ could have leaks under other channels.

\item These libraries have a large inherited codebase developed when side-channel attacks were not a concept,
therefore migrating the entire ECC stack to constant-time code, while doable, require significant effort.

\end{enumerate}

Regarding the individual responses from libraries,
\libgcrypt does not implement projective coordinates randomization.
However, as part of the disclosure, we were asked to test an upcoming constant-time implementation
for the Curve25519 field, hence we repeated the experiments performed in \autoref{sec:libgcrypt_analysis}.
We found that all page combinations generated a single leakage trace under \pt and \cc.
Therefore, according to our evaluation flow, said code can be considered OTA-safe regarding these side channels.

The \mbedtls team decided to randomize the projective coordinates \emph{before} and \emph{after}
the scalar multiplication, preventing both \emph{forward} and \emph{backward} OTAs.
This was logical, since their library already featured a randomization primitive implementation.
Finally, they fixed the bug preventing randomization in some cases
(see \autoref{sec:attack_mbedtls} for details).

As a consequence of our work, \wolfssl implemented the coordinates randomization countermeasure
to prevent both address and value-based OTAs against its ECC implementation.
In addition, their library features a very mature constant-time implementation,
enabled by setting the non-default flags \code{WOLFSSL\_SP} and \code{WOLFSSL\_HAVE\_SP\_ECC}.
Upon their request,
we tested said implementation, repeating the \autoref{sec:wolfssl_analysis} experiments
as part of the disclosure, and found no evidence of leakage under \pt and \cc side channels.
 
\end{document}